\documentclass[12pt]{article}
\usepackage{pdproc} 
\usepackage{epsfig}
\def\H{H\hskip-8.5pt/\hskip2pt}

\begin{document}

\renewcommand{\thefootnote}{\alph{footnote}}

\title{NEUTRINOS AND THE PHENOMENOLOGY OF CPT VIOLATION }

\author{N.~E.~MAVROMATOS}

\address{ Department of Physics-Theoretical Physics, 
King's College London, \\
 Strand, London, WC2R 2LS, U.K.\\
 {\rm E-mail: nikolaos.mavromatos@cern.ch}}




\abstract{In this talk I review briefly theoretical models and ideas
on quantum gravity approaches entailing CPT violation.
Then, I discuss various phenomenological tests of CPT violation
using neutrinos, arguing in favour of their superior sensitivity 
as compared to that of 
tests using other 
particles, such as neutral mesons, or nuclear and atomic physics 
experiments. I stress the fact that 
there is no single figure of merit for CPT violation, 
and that the conclusions on phenomenological
sensitivities drawn so far are highly quantum-gravity-model dependent.}
   
\normalsize\baselineskip=15pt

\section{Introduction and Summary}
  
There is a number of fundamental questions that one has to ask
before embarking on a study of the phenomenology of CPT violation:

(I) Are there theories which 
allow CPT breaking?

(II) How (un)likely is it that somebody someday 
finds CPT violation,
and why?

(III) What formalism does one has to adopt? 
How can we be sure of observing CPT 
Violation and not something else? our current phenomenology 
of particle physics is based on CPT 
invariance.

(IV) There does not seem to be a single 
``figure of merit'' for CPT violation. Then how should we 
compare various ``figures of merit" of CPT tests
(e.g. direct
mass measurement between matter and antimatter, $K^0$-${\overline K}^0$ 
mass difference a la CPLEAR, Decoherence Effects, Einstein-Podolsky-Rosen 
(EPR) states in meson factories, 
neutrino mixing, electron g-2 and 
cyclotron frequency comparison, 
neutrino spin-flavour conversion {\it etc}.)

In some of these questions we shall try to give  some answers
in the context of this presentation. Because this is a conference
on neutrinos, I will place emphasis on neutrino tests of CPT invariance.
As I will argue below, in many instances neutrinos seem to provide at present
the best bounds on possible CPT violation. However, I must stress
that, precisely because CPT violation is a highly model dependent
feature of some approaches to quantum gravity (QG), there may be 
models in which the sensitivity of other experiments on CPT violation, 
such as astrophysical experiments, is superior to that of current neutrino 
experiments.

My talk will focus on the following three major issues:  

{\bf (a)} \underline{WHAT IS CPT SYMMETRY}: I will give a definition of what we mean 
by CPT invariance, and under what conditions this invariance holds.

{\bf (b)} \underline{WHY CPT VIOLATION ?}: Currently there are various 
{\it Quantum Gravity Models} which  
may {\it violate} Lorentz symmetry and/or
quantum coherence (unitarity {\em etc}), and through this CPT symmetry: 

(i) space-time foam (local field theories, non-critical strings {\em etc.}), 

(ii) (non supersymmetric) string-inspired 
standard model extension with Lorentz Violation. 

(iii)  Loop Quantum Gravity. 

(iv) However, {\em CPT violation} may also occur at a {\em global scale}, 
{\em cosmologically}, 
as a result of a 
cosmological constant in the Universe, whose presence may jeopardize 
the definition of a standard scattering matrix.

{\bf (c)} \underline{HOW CAN WE DETECT CPT VIOLATION?} : Here is a 
current list of most sensitive 
particle physics probes for CPT tests: 
(i) {\em Neutral Mesons}: KAONS, B-MESONS, 
entangled states in $\phi$ and $B$ factories. 

(ii) {\em anti-matter factories}: antihydrogen 
(precision spectroscopic tests on free and trapped 
molecules ), 

(iii) Low energy atomic physics experiments, including ultra cold neutron
experiments in the gravitational field of the Earth. 

(iv) Astrophysical Tests  
(especially  Lorentz-Invariance
violation  tests, via  modified  dispersion relations of matter 
probes {\it etc.})

(iv) Neutrino Physics, on which we shall mainly concentrate in this talk.

I shall be brief in my description due to space restrictions. For 
more details I refer the interested reader to the relevant literature.
I have tried to be as complete as possible 
in reviewing the phenomenology of CPT violation for neutrinos, 
but I realize that I might not have done a complete job; I 
should therefore apologize for possible omissions in references, 
but this is not 
intentional. I do hope, however, that I give a satisfactory representation
of the current situation. 

\section{The CPT theorem and how it may be evaded}

The CPT theorem refers to quantum field theoretic models of particle physics,
and ensures their invariance under  the successive operation (in any order)
of
{\bf  C}(harge), {\bf P}(arity=reflection), and {\bf T}(ime reversal).
The {\em invariance} of the Lagrangian density ${\cal L}(x)$ 
of the field theory under the combined action of {\em CPT}  
is a property of any quantum field theory  in a {\em Flat} space time  
which respects:
{\em  (i) Locality, (ii) Unitarity and (iii) Lorentz Symmetry}.
\begin{eqnarray} 
\Theta {\cal L}(x)\Theta^\dagger = {\cal L}(-x)~,~\Theta =CPT~,~{\cal L}={\cal L}^\dagger 
\label{lagrcpt}
\end{eqnarray}

The theorem has been suggested first by L\"uders and Pauli~\cite{pauli}, and 
also by John Bell~\cite{bell},
and has been put on an axiomatic form, using Wightman axiomatic approach to 
relativistic (Lorentz invariant) field theory by Jost~\cite{jost}. 
Recently the Lorentz covariance of the Wightmann (correlation) functions
of field theories~\cite{wight} 
for a proof of CPT has been re-emphasized in \cite{greenb}, in 
a concise simplified exposition of the work of Jost.  
The important point to notice in that proof is the use of 
flat-space Lorentz covariance, which allows the passage onto a momentum 
(Fourier) formalism. Basically, the Fourier formalism employs 
appropriately superimposed plane 
wave solutions for fields, with four-momentum $p_\mu$. 
The proof of CPT, then, follows by the Lorentz covariance
transformation properties of the Wightman functions, and the unitarity 
of the Lorentz transformations of the various fields.

In curved space times, especially highly curved ones
with space-time boundaries, such as 
space-times in the (exterior) 
vicinity of black holes, where the boundary is provided
by the black hole horizons, 
or space-time foamy 
situations, in which one has vacuum creation of 
microscopic (of Planckian size $\ell _P = 10^{-35}$ m) black-hole 
horizons~\cite{foam}, such an approach is invalid, and Lorentz
invariance, and may be unitarity, are lost.  Hence such models of quantum 
gravity
violate (ii) \& (iii) of CPT theorem, and hence one should expect its {\em
violation}.  

It is worthy of discussing briefly the basic mechanism by which 
unitarity may be lost in 
space-time foamy situations in quantum gravity. This is the speaker's favorite
route for possible quantum-gravity induced CPT violation, which may hold 
independently of possible Lorentz invariant violations. It is at the core of 
the induced decoherence by quantum gravity~\cite{ehns,emn}.

\begin{figure}[ht]
\centering
  \epsfig{file=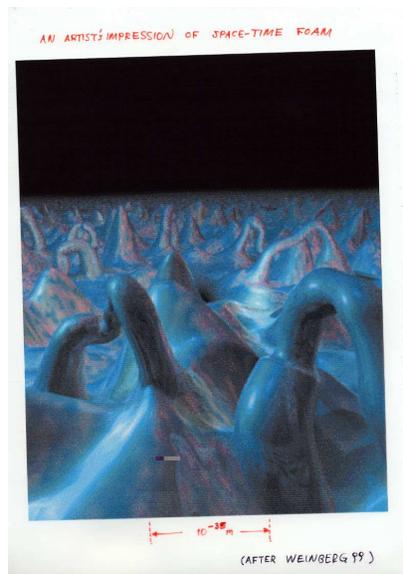, width=0.5\textwidth,bbllx=0,bblly=0,bburx=1982,bbury=1982,clip=}
\caption{An artist impression of a foamy space time (S. Weinberg, 
{\it Sci. Am., ``A Unified Physics by 2050?} (December 1999 issue))}.
\label{foamfig}
\end{figure}

The important point to notice is that in general 
space-time may be {\em discrete} and {\it topologically non-trivial} 
at Planck scales  $10^{-35}~m$ (see fig. \ref{foamfig}), 
which would in general 
imply  Lorentz symmetry Violation (LV), and hence CPT violation (CPTV). 
Phenomenologically, at a macroscopic level, such LV may lead to 
extensions of the standard model which violate both Lorentz and 
CPT invariance~\cite{kostel}.

\begin{figure}[ht]
\centering
  \epsfig{file=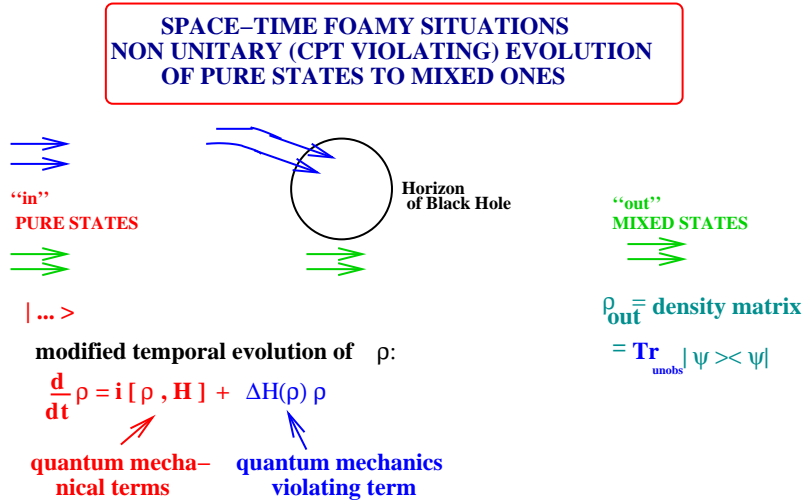, width=0.7\textwidth}
\caption{A basic mechanism for loss of information in a space time foamy
situation.}
\label{loss}
\end{figure}

In addition,  
there may be an {\it environment of gravitational} 
degrees of freedom (d.o.f.) {\it inaccessible} to 
low-energy experiments (for example non-propagating d.o.f., for which
ordinary 
scattering is not well defined~\cite{emn}). This will lead in general
to an 
{\it apparent information loss} for low-energy observers,
who by definition  can measure only propagating low-energy
d.o.f. by means of scattering experiments. 
As a consequence, an apparent lack of unitarity and hence 
CPTV may arise, which is in principle independent of any LV effects. 
The loss of information may be understood simply 
by the mechanism illustrated in fig. \ref{loss}. In a 
foamy space time there is an eternal creation and annihilation 
of Quantum Gravity {\em singular} fluctuations  (e.g. microscopic
(Planck size) black holes etc), which indeed imply that 
the observable space time is an open system. When matter particles
pass by such fluctuations (whose life time is Planckian, of order
$10^{-43}$ s),  part of 
the particle's quantum numbers ``fall into'' the horizons, and are captured
by them as the microscopic horizon disappears into the foamy vacuum.   
This may imply  the exchange of information between 
the observable world and the gravitational ``environment'' consisting 
of degrees of freedom inaccessible to low energy scattering experiments,
such as back reaction of the absorbed matter onto the space time, recoil of the microscopic black hole {\it etc.}. 
In turn, such a loss of information will imply evolution of 
initially pure quantum-mechanical states 
to mixed ones for an asymptotic observer.

As a result, the asymptotic observer will have to use density matrices
instead of pure states: 
$\rho_{out}={\rm Tr}_{\rm unobs}|out><out|=$ \$ $\rho_{in} \\
 \$ \ne S S^\dagger,$, with $S=e^{iHt}$ the ordinary 
scattering matrix. Hence, in a foamy situation the concept of the 
scattering matrix is replaced by that of 
\$, introduced by Hawking~\cite{foam}, which   
is {\it non invertible}, and in this 
way it quantifies the unitarity loss in the effective low-energy theory. 
The latter violates CPT due to a mathematical theorem by R. Wald, 
which we now describe~\cite{wald}.  

Notice that this is an effective violation,
and indeed the complete theory of quantum gravity (which though is still 
unknown) may respect some form of CPT invariance. 
However, from a phenomenological point of view, this 
effective low-energy violation of CPT is the kind of violation 
we are interested in.

\subsection{\$  matrix and CPT Violation (CPTV)}
 
The theorem states the following~\cite{wald}: 
{\em if \$ $\ne S~S^\dagger $, then CPT is violated, at least in its
strong form, in the sense that the CPT operator is not well defined.}

For instructive purposes we shall give here an elementary proof.
Suppose that CPT is conserved, then there exists a unitary, invertible 
CPT operator
$\Theta$: $\Theta {\overline \rho}_{in} = \rho_{out}.$

We have $ \rho_{out} =$ \$ $\rho_{in} \to \Theta {\overline \rho}_{in} 
=$\$ $\Theta^{-1} {\overline \rho}_{out} \to $ 
${\overline \rho}_{in} =\Theta^{-1} $\$ $\Theta^{-1} {\overline \rho}_{out} $.

But ${\overline \rho}_{out} =$\$${\overline \rho}_{in} $, 
hence : ${\overline \rho}_{in} = \Theta^{-1}$\$$\Theta^{-1} $ \$ 
 ${\overline \rho}_{in}$.

The last relation implies that \$  has an {\it inverse} 
\$$^{-1} = \Theta^{-1}$\$$\Theta^{-1} $, which however as 
we explained above is {\it impossible} due to information loss;  hence
CPT {\em must be violated} (at least in its strong form, i.e. $\Theta$ 
is not a well-defined operator).
As I remarked before this is 
my preferred way of CPTV by Quantum Gravity, given that it may occur
in general independently of LV and thus preferred frame approaches
to quantum gravity. Indeed, I should stress at this point that 
the above-mentioned gravitational-environment induced decoherence
may be Lorentz invariant~\cite{mill}, the appropriate Lorentz transformations
being slightly modified to account, for instance, for the discreteness
of space time at Planck length~\cite{discr}. 
This is an interesting 
topic for research, and it is by no means complete. 
Although the lack of an invertible scattering matrix
in most of these cases implies a strong violation of CPT, 
nevertheless, 
it is interesting to
demonstrate explicitly whether some form of CPT invariance holds 
in such cases~\cite{waldron}. This also includes cases with non-linear
modifications of Lorentz symmetry~\cite{nlls}, arising from the requirement 
of viewing the Planck length as an invariant-observer independent
proper length in space time. 

It should be stressed at this stage that, if the CPT operator 
is not well defined,
then this may lead to a whole new perspective of dealing with precision 
tests in meson factories. In the usual LV case of CPTV~\cite{kostel}, the 
CPT breaking is due to the fact that the CPT operator, which is well-defined
as a quantum mechanical operator in this case, does not commute with the 
effective low-energy Hamiltonian of the matter system. This leads to a
mass difference between particle and antiparticle. If, however, the 
CPT operator is not well defined, as is the case of the quantum-gravity
induced decoherence~\cite{ehns,emn}, 
then, the concept of the `antiparticle' gets 
modified~\cite{bernabeu}. In particular, the antiparticle space is 
viewed as an independent subspace of the state space of the system,
implying that, in the case of neutral mesons, for instance, the 
anti-neutral meson should not be treated as an identical particle
with the corresponding meson. This leads to the possibility of novel 
effects associated with CPTV as regards EPR states, which may be testable
at meson factories~\cite{bernabeu}. 

Another reason why I prefer the CPTV via the \$ matrix decoherence
approach concerns a novel type of CPT violation {\it at a global scale}, 
which may characterize
our Universe, what I would call 
{\it cosmological CPT Violation}, proposed in ref.~\cite{cosmonem}.

\subsection{Cosmological CPTV?}

\begin{figure}[htb]
  \centering
\epsfig{file=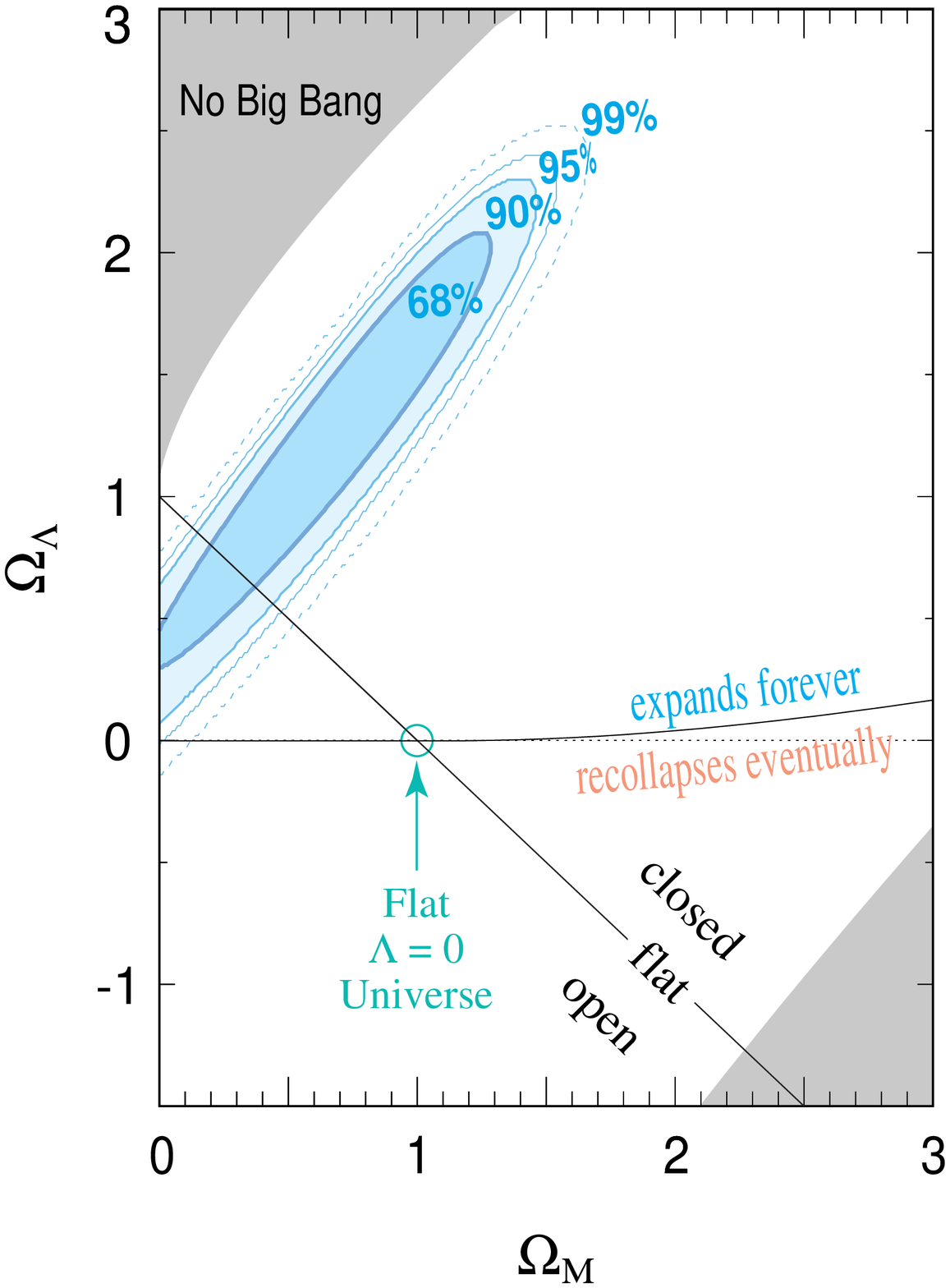, width=0.4\textwidth}
\hfill 
\epsfig{file=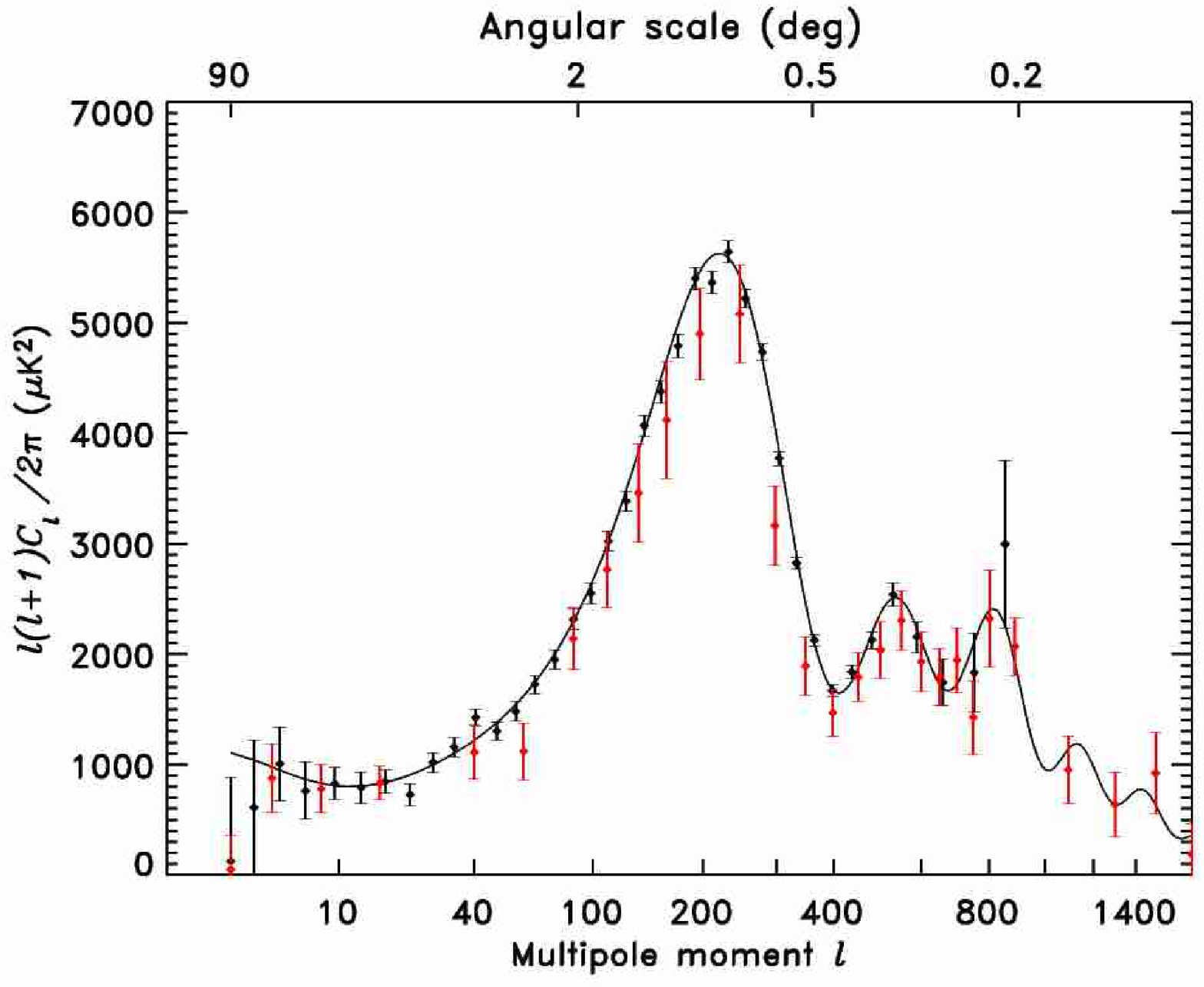,angle=90,width=0.4\textwidth}
\hfill 
  \epsfig{file=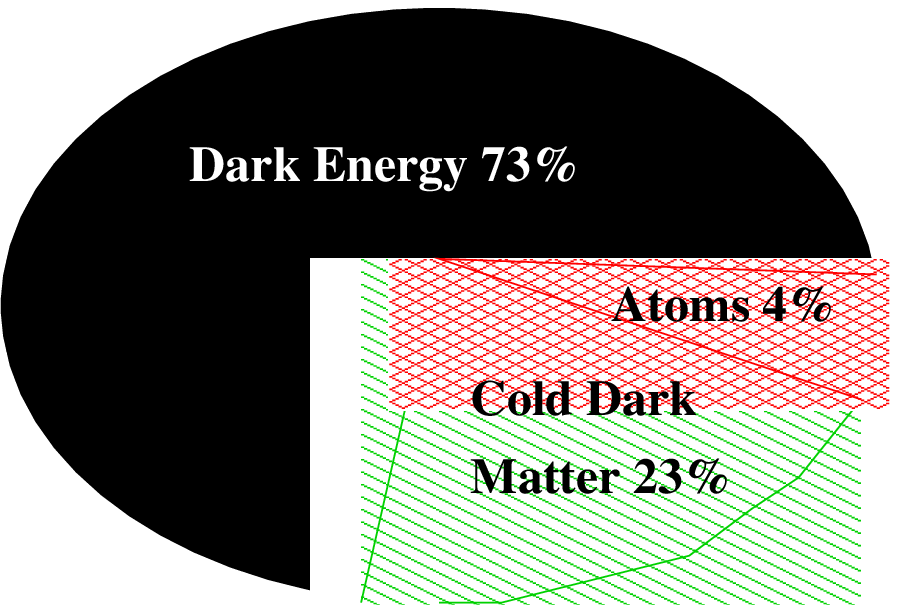, width=0.4\textwidth}
\caption{Recent observational evidence for Dark Energy of the Universe 
({\em upper left figure}: evidence from SnIa (Ref. [17]), {\em upper right figure}: 
evidence from CMB measurements (Ref. [18])) 
and a pie graph ({\em lower central figure}) of the energy budget of our world 
according to these observations.}
\end{figure}

This type of CPTV is prompted by recent 
astrophysical Evidence for the existence of a Dark Energy 
component of the Universe. For instance, there is direct evidence 
for a current era acceleration of the Universe, based on measurements
of distant supernovae SnIa~\cite{snIa}, which is supported also by 
complementary observations on Cosmic Microwave Background (CMB) 
anisotropies (most spectacularly by the recent data of WMAP satellite 
experiment)~\cite{wmap}. 

Best fit models of the Universe from such combined data 
appear at presence consistent with 
a non-zero {\em cosmological constant} $\Lambda \ne 0$. 
Such a $\Lambda$-universe will eternally accelerate, as it will enter 
eventually an inflationary (de Sitter) phase again, in which the scale 
factor will diverge exponentially   
$a(t) \sim e^{\sqrt{\Lambda/3}t}$,
$t \to \infty$. 
This implies that there 
exists a {\bf cosmological Horizon}. 

The existence of such Horizons implies incompatibility with 
a S-matrix: 
no proper definition of asymptotic state vectors is possible, 
 and there is always an 
environment of d.o.f. crossing the horizon.  
This situation  may be considered as dual to 
that of black hole, depicted in fig. \ref{loss}: in that case the 
asymptotic observer was in the exterior of the black hole horizon,  
in the cosmological case the observer is inside the horizon.
However, both situations are characterized 
by a lack of an invertible scattering matrix, hence the above-described 
theorem by Wald on \$-matrix and CPTV applies~\cite{cosmonem}, 
and thus CPT is violated
due to a cosmological constant $\Lambda >0 $. It has been argued in 
\cite{cosmonem} that such a violation is described effective by a 
modified temporal evolution of matter in such a $\Lambda$-universe, 
which is given by 
\begin{equation}  
\partial_t \rho = [\rho, H] +  {\cal O}(\Lambda/M_P^3)\rho 
\label{densityevol}
\end{equation} 
Notice that although the order of the
cosmological CPTV effects is tiny, 
if we accept that the Planck scale is the ordinary four-dimensional one 
$M_P \sim 10^{-19}$ GeV, and hence undetectable in direct particle
physics interactions, 
however, as we have seen from the above considerations, they may have already 
been  
detected indirectly through the (claimed) observational 
evidence for a current-era acceleration of the 
Universe! Of course, the existence of a cosmological constant brings up 
other interesting challenges, such as the possibility of  
a proper quantization of de Sitter space as an open system, 
which are still unsolved.

At this point I should mention that 
time Relaxation models 
for Dark Energy (e.g. quintessence models), where
eventually the vacuum energy asymptotes (in cosmological time) 
an equilibrium zero value are still currently 
compatible with the data~\cite{relax}. In such cases it might be possible that 
there is no cosmological CPTV, since a proper S-matrix can be defined,
due to lack of cosmological horizons. 

From the point of view of string theory the impossibility of defining a 
S-matrix is very problematic, because critical strings by their very definition
depend crucially on such a concept. However, this is not the case of 
non-critical (a kind of non equilibrium) string theory, which can accommodate
in their formalism $\Lambda$ universes~\cite{cosmonem}.
It is worthy of mentioning briefly that such non-critical (non-equilibrium) 
string theory cases are capable of accommodating models with large 
extra dimensions, in which the string gravitational scale $M_s$ is not 
necessarily the same as the Planck scale $M_P$, but it could be much 
smaller, e.g. in the range of a few TeV.
In such cases, the CPTV effects in (\ref{densityevol}) may be much larger,
since they would be suppressed by $M_s$ rather than $M_P$, and also they will
be proportional to compactification volumes of the (large) extra dimensions. 

It would be interesting to study further the cosmology of such models
and see whether the global type of CPTV proposed in \cite{cosmonem}, 
which also entails primordial CP violation of similar order, 
distinct from the ordinary (observed) CP violation which 
occurs at a later stage in the evolution of the Early Universe, 
may provide a realistic explanation of the initial matter-antimatter
asymmetry in the Universe, and the fact that antimatter is 
highly suppressed today. In the standard CPT invariant 
approach this asymmetry is supposed to be 
due to ordinary CP violation. In this respect, I mention at this point that 
speculations about the possibility that 
a primordial CPTV space-time foam is responsible
for the observed matter-antimatter asymmetry in the Universe 
have also been put forward in \cite{ahluwal} but from a different
perspective than the one I am suggesting here. In ref. \cite{ahluwal}  
it was suggested that a novel CPTV foam-induced phase difference 
between a space-time spinor and its antiparticle 
may be responsible for the required asymmetry. 
Similar properties of spinors may also characterize space times 
with deformed 
Poincare symmetries~\cite{agostini}, which may also be viewed as 
candidate models of quantum gravity.
In addition, other attempts to discuss the origin of such an asymmetry 
in the Universe 
have been made within the 
loop gravity approach to quantum gravity~\cite{singh} 
exploring Lorentz Violating 
modified dispersion relations for matter probes, especially neutrinos, 
which we shall discuss below.

\section{Phenomenology of CPT Violation} 

\subsection{Order of Magnitude Estimates of CPTV} 

Before embarking on a detailed phenomenology of CPTV it is worth asking whether such a task is really sensible, in other words how feasible it is to detect 
such effects in
the foreseeable future. 
To answer this question
we should present some estimates of the expected effects in some 
models of quantum gravity. 

The order of magnitude of the CPTV effects 
is a highly model dependent issue, and it depends crucially on the specific way CPT is violated in a
model. As we have seen  
cosmological (global) CPTV effects are tiny, 
on the other hand, quantum Gravity   
(local) space-time effects (e.g. space time foam) may be much larger
for the following reason:  
Naively, Quantum Gravity (QG) has a dimensionful constant:
$G_N \sim 1/M_P^2$, $M_P =10^{19}$ GeV. Hence, CPT violating 
and decoherening 
effects may be expected to be suppressed
by  $E^3/M_P^2 $ , where $E$ is a typical energy scale of the low-energy 
probe.  This would be practically undetectable in neutral mesons,
but some neutrino 
flavour-oscillation experiments (in models where flavour symmetry
is broken by quantum gravity), or some 
cosmic neutrino future observations 
might be sensitive to this order: 
for instance, in models with LV, one expects 
modified dispersion relations (m.d.r.) which could yield significant effects
for ultrahigh energy ($10^{19}$ eV)
$\nu$ from Gamma Ray Bursts (GRB)~\cite{volkov}, 
that could be close to observation. 
Also in some astrophysical cases, e.g. 
observations of 
synchrotron radiation 
Crab Nebula or Vela pulsar, one is able to  
constraint electron m.d.r. almost near this 
(quadratic) order~\cite{synchro}. 

However, resummation and other effects 
in some theoretical 
models may result in much larger CPTV effects of order: 
 $\frac{E^2}{M_P}$. This happens, e.g., in some loop gravity 
models~\cite{loop}, or in 
some (non-critical) stringy models of quantum gravity involving 
open string excitations~\cite{emn}.
Such large effects are already accessible in current experiments,
and as we shall see most of them are excluded by current observations. 
The Crab nebula synchrotron constraint~\cite{crab} for instance already 
excludes such effects
for electrons.

\subsection{Mnemonic Cubes for CPTV Phenomenology}

\begin{figure}[htb]
\centering
  \epsfig{file=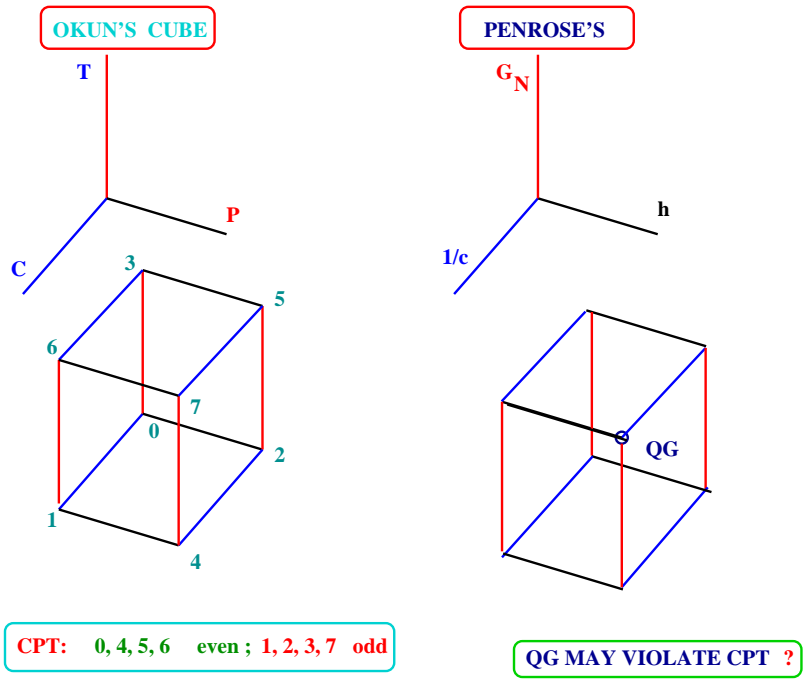, width=0.7\textwidth}
\caption{Mnemonic cubes for CPT Violation: \underline{Left} its phenomenology.
\underline{Right}: its possible theoretical origin.}
\label{mnemonic}
\end{figure}

When CPT is violated there are many possibilities, due to the fact that 
C,P and T may be violated individually, with their violation 
independent from one another.
This was emphasized by Okun~\cite{okun} some years ago, who presented a set of 
mnemonic rules for CPTV phenomenology, which are summarized in 
fig.~\ref{mnemonic}.
In this figure I also draw a sort of 
Penrose cube, indicating where the violations 
of CPT may come from (the diagram has to be interpreted as follows: CPTV may come 
from violations of special relativity (axis $1/c$), 
where the speed of light does not have its value, 
having some sort of refractive index, 
from departure of quantum mechanics (axis $h$), from gravity considerations
where the gravitational constant departs from its value (axis $G_N$), 
or finally (and most likely)  
from quantum gravity considerations  where all such effects may 
coexist.

\subsection{Lorentz Violation and CPT: The Standard Model Extension (SME)} 

We start our discussion on phenomenology of CPT violation by 
considering CPTV models 
in which requirement (iii) of the CPT theorem is violated, that of Lorentz invariance. As mentioned previously, such a violation may be a consequence
of quantum gravity fluctuations. 
In this case Lorentz symmetry is violated and hence CPT,
but there is no necessarily quantum decoherence or unitarity loss. 
Phenomenologically, at low energies, 
such a LV will manifest itself as an extension of the standard 
model in (effectively) flat space times, whereby LV terms will be introduced
by hand in the relevant lagrangian, with coefficients whose magnitude
will be bounded by experiment~\cite{kostel}. 

Such SME lagrangians may be viewed as the low energy limit of 
string theory vacua, in which  
some tensorial fields acquire non-trivial vacuum expectation values 
$<A_\mu>  \ne 0~, <T_{\mu_1 \dots \mu_n}>  \ne 0.$
This implies a 
{\bf spontaneous} breaking of Lorentz symmetry by  
these (exotic)  string vacua~\cite{kostel}.

The simplest phenomenology of CPTV in the context of SME 
is done by studying the physical consequences of a modified Dirac equation 
for charged fermion fields in SME. 
This is relevant for phenomenology using data from the recently 
produced antihydrogen factories~\cite{antihydro,mavroyoko}. 

In this talk I will not cover this part in detail, 
as I will concentrate mainly 
in neutrinos within the SME context. It suffices to mention that 
for free hydrogen  
$H$  (anti-hydrogen ${\overline H}$) one may consider 
the spinor $\psi$  representing electron  
(positron)  with charge $q=-|e| (q=|e|)$ 
around a proton (antiproton)   
of charge $-q$, which obeys the modified Dirac equation (MDE):   
\begin{eqnarray}
&& ( i\gamma^\mu D^\mu - M -  
a_\mu \gamma^\mu - b_\mu \gamma_5 \gamma^\mu - \nonumber \\
&& -\frac{1}{2}H_{\mu\nu}\sigma ^{\mu\nu}
+  ic_{\mu\nu}\gamma^\mu D^\nu + id_{\mu\nu}\gamma_5\gamma^\mu D^\nu  )\psi =0 \label{esem}
\end{eqnarray} 
where $D_\mu = \partial_\mu - q A_\mu$, $A_\mu = (-q/4\pi r, 0)$ Coulomb 
potential. CPT \& Lorentz violation is described by terms
with parameters $a_\mu~, b_\mu~,$ while 
Lorentz violation only is described by the terms
with coefficients $c_{\mu\nu}~, d_{\mu\nu}~,  H_{\mu\nu} $.  

One can perform spectroscopic tests on free and magnetically trapped molecules,
looking essentially for transitions that were forbidden if CPTV and SME/MDE 
were not taking place. 
The basic conclusion is that for sensitive tests of CPT in antimatter 
factories frequency resolution in spectroscopic measurements has to 
be improved down to a range of 
a 1 mHz, which at present is far from being achieved~\cite{mavroyoko}.

\begin{figure}
\centering
\epsfig{file=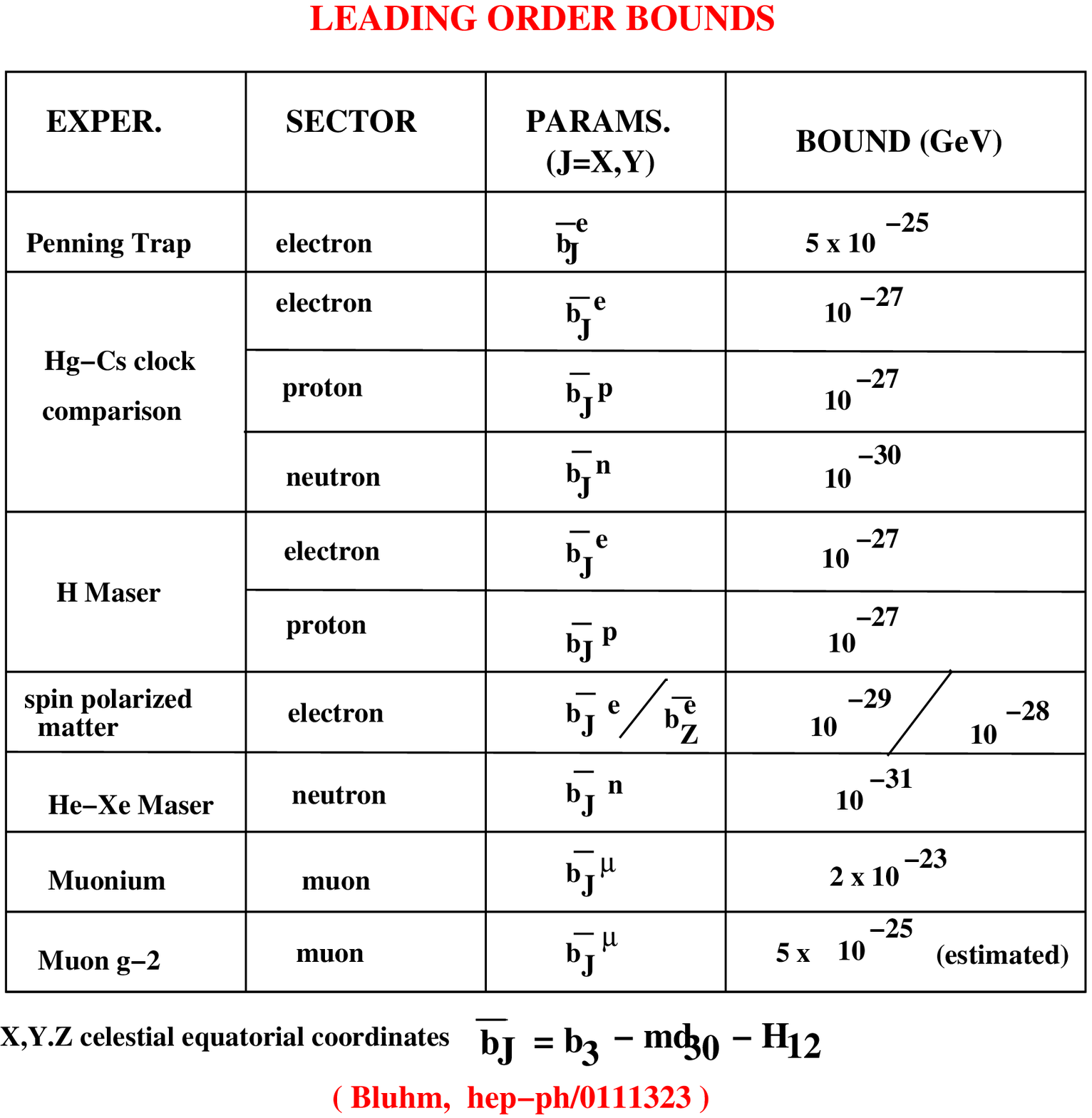, width=0.6\textwidth}
\caption{Sensitivities of CPTV and LV parameters appearing in 
SME/Modified Dirac equation for charged probes,  
from various atomic and nuclear physics experiments}
\label{bluhm}
\end{figure}

Since the presence of LV interactions in the SME affects dispersion relations
of matter probes, other interesting precision tests of such extensions
can be made in atomic and nuclear physics experiments, exploiting the fact
of the existence of a preferred frame where observations take place. 
The results and the respective sensitivities of the various parameters
appearing in SME are summarized in the table of figure \ref{bluhm},
taken from the paper of \cite{bluhm}. 

\subsection{Direct SME Tests for Neutrinos and 
Modified Dispersion relations (MDR)}

Many LV Models of Quantum Gravity (QG) predict modified dispersion 
relations (MDR) for matter probes, including  
neutrinos $\nu$~\cite{emnnature,volkov,grb}.
This leads to one important class of experimental tests using $\nu$: 
each mass eigenstate of $\nu$ 
has QG deformed dispersion relations, which 
may, or may not, be the same for all flavours: 
\begin{eqnarray} 
E^2 = {\vec k}^2 + m_i^2 + f_i(E,M_{qg},{\vec k})~, \qquad 
{\rm e.g.}~ 
f_i = \sum_\alpha C_{\alpha=1,2,...} {\vec k}^2 (\frac{|{\vec k}|}{M_{qg}})^\alpha
\label{mdrnu}
\end{eqnarray}
There are stringent bounds on $f_i$ from oscillation experiments, as 
we shall discuss below. 

It must be stressed that such MDR also characterize SME, although the 
origin of MDR in the 
approach of \cite{emnnature,volkov,grb} is due to an induced 
non-trivial microscopic curvature of space time as a result of a 
back reaction of matter interacting with a stringy space time foam vacuum.
This is to be contrasted with 
the SME approach~\cite{kostel}, 
where the analysis is done 
exclusively on flat Minkowski space times, at a
phenomenological 
level. 

In general there are various experimental tests that can set bounds
on MDR parameters, which can be summarized as follows:

 (i) astrophysics tests - arrival time fluctuations for photons 
(model independent analysis of astrophysical GRB data~\cite{grb}

(ii) Nuclear/Atomic Physics precision measurements (clock comparison, 
spectroscopic tests on free and trapped 
molecules, quadrupole moments {\it etc})~\cite{bluhm}.

(iii) antihydrogen (precision spectroscopic tests 
on free and trapped 
molecules: e.g. $1S \to 2S$ forbidden transitions)~\cite{mavroyoko}, 

(iv) Neutrino mixing and spin-flavour conversion, 
a brief discussion of which we now turn to.

\subsection{Neutrinos and SME }

The SME formalism naturally includes the neutrino sector. 
Recently a 
SME-LV+CPTV phenomenological model for neutrinos has been given 
in \cite{mewes}. The pertinent lagrangian terms are given by: 
\begin{eqnarray} 
{\cal L}^\nu_{SME} \ni 
\frac{1}{2}i{\overline \psi}_{a,L} \gamma^\mu D_\mu \psi_{a,L}  
- (a_L)_{\mu ab}  
{\overline \psi}_{a,L} \gamma^\mu \psi_{b,L}  
+ \frac{1}{2}i  (c_L)_{\mu\nu ab}
 {\overline \psi}_{a,L} \gamma^\mu D^\nu \psi_{b,L}
\label{smenulagr}
\end{eqnarray} 
where $a,b$ are flavour indices. The model has (for simplicity)  
no $\nu$-mass differences. 
Notice that the presence of LV induces directional dependence
(sidereal effects)!.

To analyze the physical consequences of the model, one passes to an 
Effective Hamiltonian~\cite{mewes} 
\begin{eqnarray} 
(H_{\rm eff})_{ab} = |{\vec p}|\delta_{ab} + \frac{1}{|{\vec p}|}
( (a_L)^\mu  p_\mu  -  (c_L)^{\mu\nu}
p_{\mu}p_{\nu})_{ab} 
\label{effham} 
\end{eqnarray} 
Notice that $\nu$ oscillations are now controlled by
the (dimensionless) quantities $a_L L$ \& $c_L L E$ 
where L is the oscillation length. This is to be contrasted 
with the conventional case, where the relevant parameter
is associated necessarily with a $\nu$-mass difference $\Delta m$: 
$\Delta m^2 L/E$ 

There is an important feature of the SME/$\nu$: 
despite CPTV, the oscillation probabilities are the same between 
$\nu$ and their antiparticles, if there are no mass differences
between $\nu$ and ${\bar \nu}$:   
$P_{\nu_x \to \nu_y} = P_{{\bar \nu_x} \to {\bar \nu_y}}$. 

Experimentally, it is possible to 
bound LV+CPTV SME parameters in the neutrino sector 
with high sensitivity, if we use data 
from high energy long baseline experiments~\cite{mewes}. Indeed, 
from the fact that 
there is   
no evidence for $\nu_{e,\mu} \to \nu_{\tau}$
oscillations, for instance, 
at $E \sim 100$ GeV ,  $L \sim 10^{-18}$ 
GeV$^{-1}$  we conclude that 
$a_L < 10^{-18}$ GeV, $c_L < 10^{-20}$. 

Similarly for an explanation of the LSND anomaly~\cite{lsnd}, 
claiming evidence for oscillations 
between antineutrinos (${\bar \nu_\mu}-{\bar \nu_e}$) 
but not for the corresponding neutrinos, a mass-squared difference of order 
 $\Delta m^2 = 10^{-19} $ GeV$^2=10^{-1}$ eV$^2$
is required, which implies that  
$a_L \sim 10^{-18}$ GeV,
$c_L \sim 10^{-17}$. This would affect other experiments, and in fact 
one can easily come to the conclusion that SME/$\nu$ does not offer a 
good explanation for LSND, if we accept the result of that experiment
as correct, which is not clear at present.   

\begin{figure}
\centering
  \epsfig{file=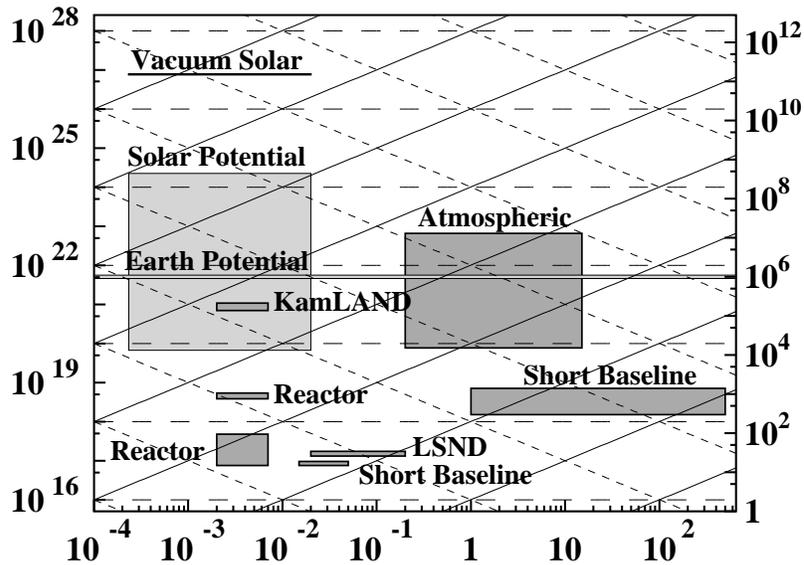, width=0.7\textwidth}
\caption{Approximate experimental sensitivities (from Ref. [33]).
Lines of constant $L/E$ (solid),
$L$ (dashed), and $LE$ (dotted) are shown,
which give sensitivities to $\Delta m^2$, $a_L$, and $c_L$,
respectively.}
\label{tablenu}
\end{figure}

A summary of the Experimental Sensitivities for $\nu$'s SME 
parameters are given in the 
table of figure \ref{tablenu}, taken from \cite{mewes}.

\subsection{Lorentz non-invariance, MDR and $\nu$-oscillations}

Models of quantum gravity predicting MDR of the type (\ref{mdrnu})
for neutrinos~\cite{volkov,alfaro}, with a leading order $E^2/M_{qg}$
modification,  
can be 
severely constrained by a study of the induced oscillations 
between neutrino flavours, as a result of the departure from 
the standard dispersion relations provided that 
the quantum-gravity foam responsible for the MDR breaks
flavour symmetry, which however is not always the case~\cite{emnnu}.

\begin{figure}
\centering
  \epsfig{file=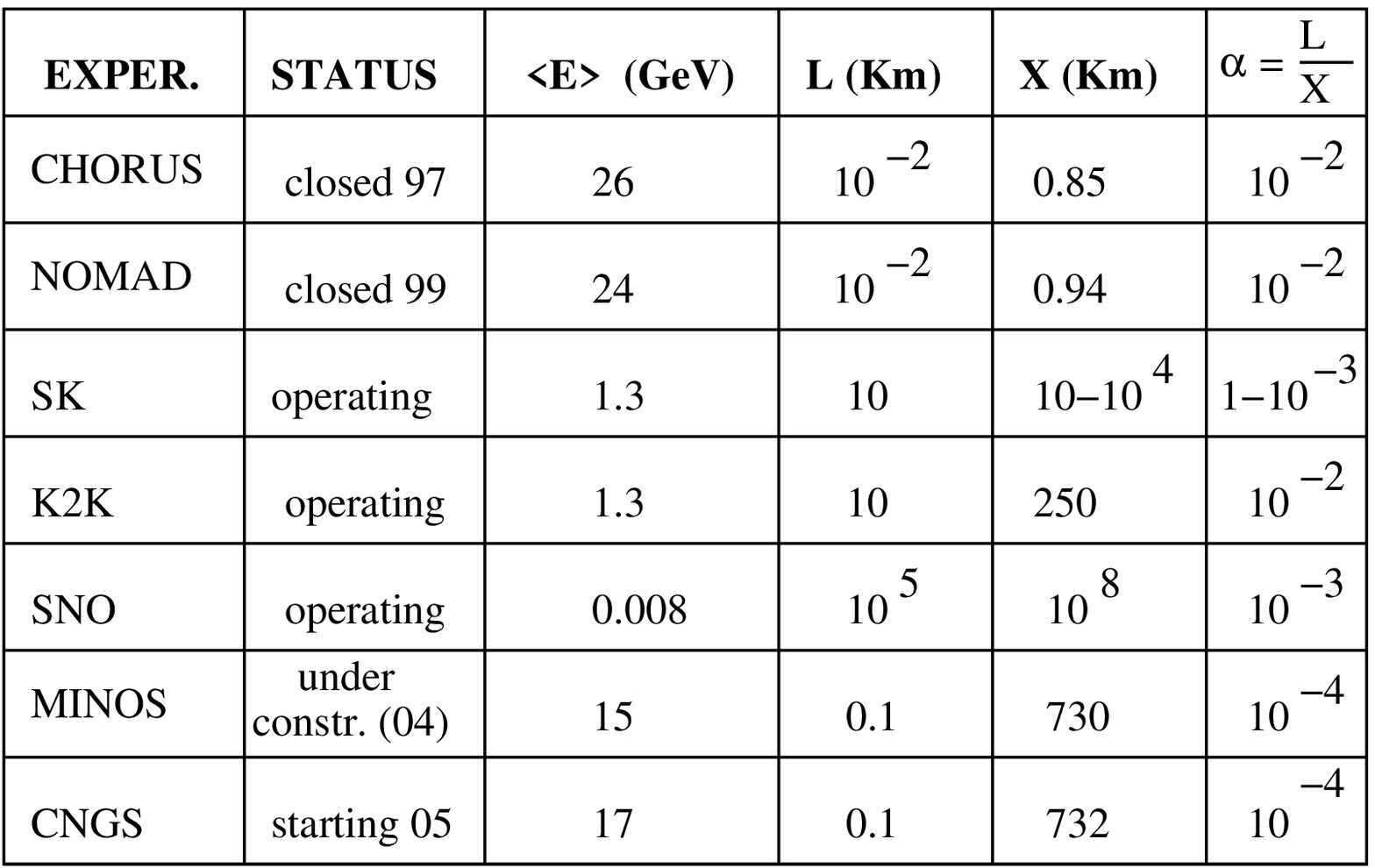, width=0.7\textwidth}
\caption{Shown for each experiment are (from Ref. [37]): (i) operation status, 
(ii) mean value of observed $\nu$ energy, (iii)  
the oscillation length $L$, (iv) typical $\nu$-flight distance $X$, 
and the ratio $\alpha = L/X$, which, in models where the foam 
induces $\nu$ flavour oscillations, coincides 
with the phenomenological parameter that controls the size of MDR effects.}
\label{brustein}
\end{figure}

This approach has been 
followed in \cite{eichler}, where it was shown that if flavour symmetry
is not protected in such MDR models, then the 
extra terms in (\ref{mdrnu}), proportional to $E^2/M_{qg}$ 
will  
induce an oscillation 
length $L \sim  2\pi M_{qg}/(\alpha E^2)$, 
where $\alpha$ is a phenomenological 
parameter that controls the size of the effect. 
This should be contrasted 
to the Lorentz Invariant case where $L_{\rm LI} \sim 4\pi E/\Delta m^2$,
with $\Delta m^2 $ the square mass difference between neutrino flavours.
From a field theoretic view point, terms in MDR proportional to
some positive integer power of $E^2/M_{qg} $ may behave as non-renormalizable 
operators, for instance, 
dimension five~\cite{myers} in the case of leading order 
QG effects suppressed only by a single power of $M_{qg}$. 

The sensitivity  
of the various neutrino oscillation experiments to the parameter
$\alpha$ is shown in figure \ref{brustein}~\cite{eichler}.  
The conclusion from such analyses, therefore, is that, 
if the flavour number symmetry 
is not protected in such MDR foam models, then 
neutrino observatories and long base-line experiments should have 
already observed such oscillations. As remarked above, however,
not all foam models that lead to such MDR predict such 
oscillations~\cite{emnnu}, 
and hence such constraints are highly foam-model dependent.

\subsection{Lorentz Non Invariance, MDR and $\nu$ spin-flavor conversion}

An interesting consequence of MDR in LV quantum gravity theories is 
associated with modifications to the well-known phenomenon of 
spin-flavour conversion in $\nu$ interactions~\cite{lambiase}. 
To be specific, we shall consider an example of a MDR for $\nu$ 
provided by a Loop Gravity approach to quantum gravity.
According to such an approach, the dispersion relations for neutrinos 
are modified to~\cite{alfaro}:
\begin{eqnarray}  
E^2_{\pm} =  A_p^2  p^2 +  \eta  p^4 \pm 2\lambda  p + m^2 
\label{lgmdrnu}
\end{eqnarray} 
where $A_p = 1 + \kappa_1\frac{\ell_P}{{\cal L}}, ~\eta = \kappa_3\ell^2_P, ~\lambda =  \kappa_5 \frac{\ell_P}{2{\cal L}^2}$, 
and ${\cal L}$ is a characteristic scale 
of the problem, which can be either (i) $ {\cal L} \sim E^{-1} $, or (ii) ${\cal L}$=constant.

It has been noted in \cite{lambiase} that such a modification in the 
dispersion relation will affect the form of the spin-flavour conversion
mechanism. Indeed,  
it is well known through the Mikheyev-Smirnov-Wolfenstein (MSW) 
effect~\cite{msw} that 
Weak interaction Effects of $\nu$ propagating in a medium 
result in an 
energy shift $\sqrt{2}G_F(2n_e - n_n)$, where $n_e (n_n)$'s denote 
electron (neutron) densities. In addition to such effects, 
one should also take into account 
the interaction of $\nu$ with external magnetic 
fields, $B$,
via a radiatively induced magnetic moment $\mu$, corresponding to 
a term in the effective lagrangian: 
${\cal L}_{\rm int} = \mu {\overline \psi}\sigma^{\mu\nu}F_{\mu\nu}\psi$,
with $\psi $ the neutrino fermionic field. 

According to the standard theory, the equation for evolution describing 
the {\it spin-flavour} conversion phenomenon due to 
the above-described medium and magnetic moment effects for, say, two neutrino
flavours ($\nu_e, \nu_\mu$) is given by: 
\begin{eqnarray} 
i\frac{d}{dr} \left(\begin{array}{c} \nu_{e L} 
\\\nu_{\mu L}\\\nu_{e R}\\\nu_{\mu R}\end{array}\right) = {\cal H}\left(\begin{array}{c} \nu_{e L} \\\nu_{\mu L}\\\nu_{e R}\\\nu_{\mu R}\end{array}\right)~,
\label{spinflcon} 
\end{eqnarray}
where the effective Hamiltonian  
${\cal H}$ should be corrected in the loop gravity case 
to take into account $\lambda$-effects, associated with MDR 
(\ref{lgmdrnu}) (we should notice at this stage that 
the above formalism refers to Dirac $\nu$; for Majorana $\nu$ 
one should replace: $\nu_{i L} \to \nu_i$, $\nu_{i R}
\to {\overline \nu}_i$). Details can be found in \cite{lambiase}. 

For our purposes we note that 
the Resonant Conditions for Flavour-Spin-flip are~\cite{lambiase}:
\begin{eqnarray} 
&& \nu_{e L} \to \nu_{\mu R}:  \quad 
2\lambda + \frac{\Delta m^2}{2p}{\rm cos}2\theta - \sqrt{2}G_Fn_e(r_{res}) =0
\nonumber \\
&& \nu_{\mu L} \to \nu_{e R}: \quad 
2\lambda - \frac{\Delta m^2}{2p}{\rm cos}2\theta - \sqrt{2}G_Fn_e(r_{res}) =0~.
\label{resonant}
\end{eqnarray} 
One can use above conditions to obtain bounds for $\lambda,\kappa_i$
via the oscillation probabilities for spin-flavour conversion:
\begin{eqnarray} 
P_{\nu_{eL} \to \nu_{\mu R}}=\frac{1}{2}(1 - 
{\rm cos}2{\tilde \theta}{\rm cos}2\theta)~,
\label{probsfcon}
\end{eqnarray} 
where ${\rm tan}2{\tilde \theta}(r)=\frac{4\mu B(r)E}{|\Delta m^2|{\rm cos}2\theta - 
4E\lambda  + 2\sqrt{2}G_FEn_e(r)}$.

To obtain these bounds the author of \cite{lambiase} made 
the following physically relevant assumptions: 
(a) Reasonable profiles for solar $n_e \sim n_0 e^{-10.5r/R_\odot}$, 
$n_0=85N_A {\rm cm}^{-3}$. (b) Also: $\mu \sim 10^{-11}\mu_B$.  
Then, an 
upper bound on $\lambda$ is obtained of order: 
$\lambda \le \frac{1}{2}\left(10^{-12}e^{-10.5r_{res}/R_\odot}{\rm eV} +
\frac{|\Delta m^2|}{2E}\right)$. 

To obtain bounds on $\kappa$ we need to distinguish two cases: 

{\bf (I)}  \underline{${\cal L}$=universal constant}: 
In this case, we already know from photon
dispersion tests on GRB and Active 
Galactic Nuclei (AGN)~\cite{grb,alfaro} that   
 ${\cal L} \sim 10^{-18}$ eV$^{-1}$. 
Then, from best-fit solar $\nu$-oscillations induced by MSW, 
one may use experimental values of $\Delta m^2$, 
${\rm sin}^22\theta$, and obtain the following bound on 
$\kappa_i$:  $\kappa_5 < 10^{-25}$. 
From atmospheric oscillations, in particular LSND experiment~\cite{lsnd},
$\nu_\mu \to \nu_e$ fits the data with: $|\Delta m^2| \sim eV^2$,
${\rm sin}^22\theta \sim (0.2 - 3)\times 10^{-3}$, $E_{\rm max} \sim
10$ MeV, then  $\kappa_5 < 10^{-17}$. 

{\bf (II)} \underline{${\cal L} \sim p^{-1}$  a mobile  scale}: 
In that case,  
from SOLAR oscillations, with $p \sim 1-10 {\rm MeV}$ 
one gets $\kappa_5 ={\cal O}(1-100)$, which is a natural  
range of values from a quantum-gravity view point. 
From atmospheric oscillations, for the 
maximum $\nu$ $E \sim 10$ MeV, and ${\cal L} \sim E^{-1}$, one obtains 
 $\kappa_5 \sim 10^4$, which is a very weak bound. 

The conclusion from these considerations, therefore, is that the 
experimental data seem to 
favour case {\bf (II)}, at least from a naturalness point of view.

\subsection{$\nu$-flavour states and modified Lorentz Invariance (MLI)}

An interesting recent idea~\cite{blama}, 
which we would like to discuss now briefly, 
arises from the observation of the  
peculiar way in which flavour $\nu$ states experience 
Lorentz Invariance. 
Indeed, neutrino flavour states 
are  {\it a superposition of mass eigenstates} 
 with standard dispersion relations of {\it different 
mass}. If one computes the expectation value 
of the Hamiltonian with respect to flavour states, e.g. in a 
simplified two-flavour scenario discussed in \cite{blama},
then one finds: 
\begin{eqnarray} 
E_e &=& <\nu_e|H|\nu_e> = \omega_{k,1}{\rm cos}^2\theta + \omega_{k,2}{\rm sin}^2\theta~, \nonumber \\
E_\mu &=& <\nu_\mu|H|\nu_\mu> = \omega_{k,2}{\rm cos}^2\theta + \omega_{k,1}{\rm sin}^2\theta~, 
\label{energystates}
\end{eqnarray} 
with $\theta$ the mixing angle. 

One has: $H|\nu_i>=\omega_i |\nu_i>$, $i=1,2$, where the 
$\omega_{k,i} = \sqrt{{\vec k}^2 + m_i^2}$ is a standard dispersion relation.
However, since 
the sum of two square roots in not in general a square root, 
one concludes that flavour states do not satisfy the 
standard dispersion relations. 
In general this poses a problem, as it would naively imply 
the introduction of a preferred frame, due to an apparent 
violation of the standard
linear Lorentz symmetry.

The idea of \cite{blama}, whose validity of course remains to be seen,
but which I find rather 
intriguing, and this is why I decided to include it in this review, 
is to avoid using preferred frames by  introducing instead 
non-linearly  
modified Lorentz transformations to account for the modified dispersion 
relations of the flavour states. The idea is formally similar, but physically
very different, to the approach of \cite{nlls}, in which, in order to 
ensure observer independence of the Planck length, viewed as an ordinary 
length in quantum gravity, and not as a universal coupling constant, 
one has to modify non linearly the Lorentz transformations.
The result is that flavour states satisfy the following MDR: 
\begin{equation}
E_i^2 f_i^2(E_i) - {\vec k}^2 g_i^2(E_i) = M_i^2 \quad i=e,\mu
\label{mdrflavour}
\end{equation}
One can 
determine~\cite{blama} the 
$f_i(E_i,\theta, m_i),g_i(E_i,\theta,m_i), M_i(m_i,\theta)$ 
by comparing with 
$E_i=E_(\omega_i,m_i)$ above ((c.f. (\ref{energystates})). 

Then, in the spirit of \cite{nlls}, one can 
identify the non-linear Lorentz transformation that leaves the MDR
(\ref{mdrflavour})   
invariant:  $U \circ (E, {\vec k}) = (Ef,{\vec k}g)$.

\begin{figure}
\centering
  \epsfig{file=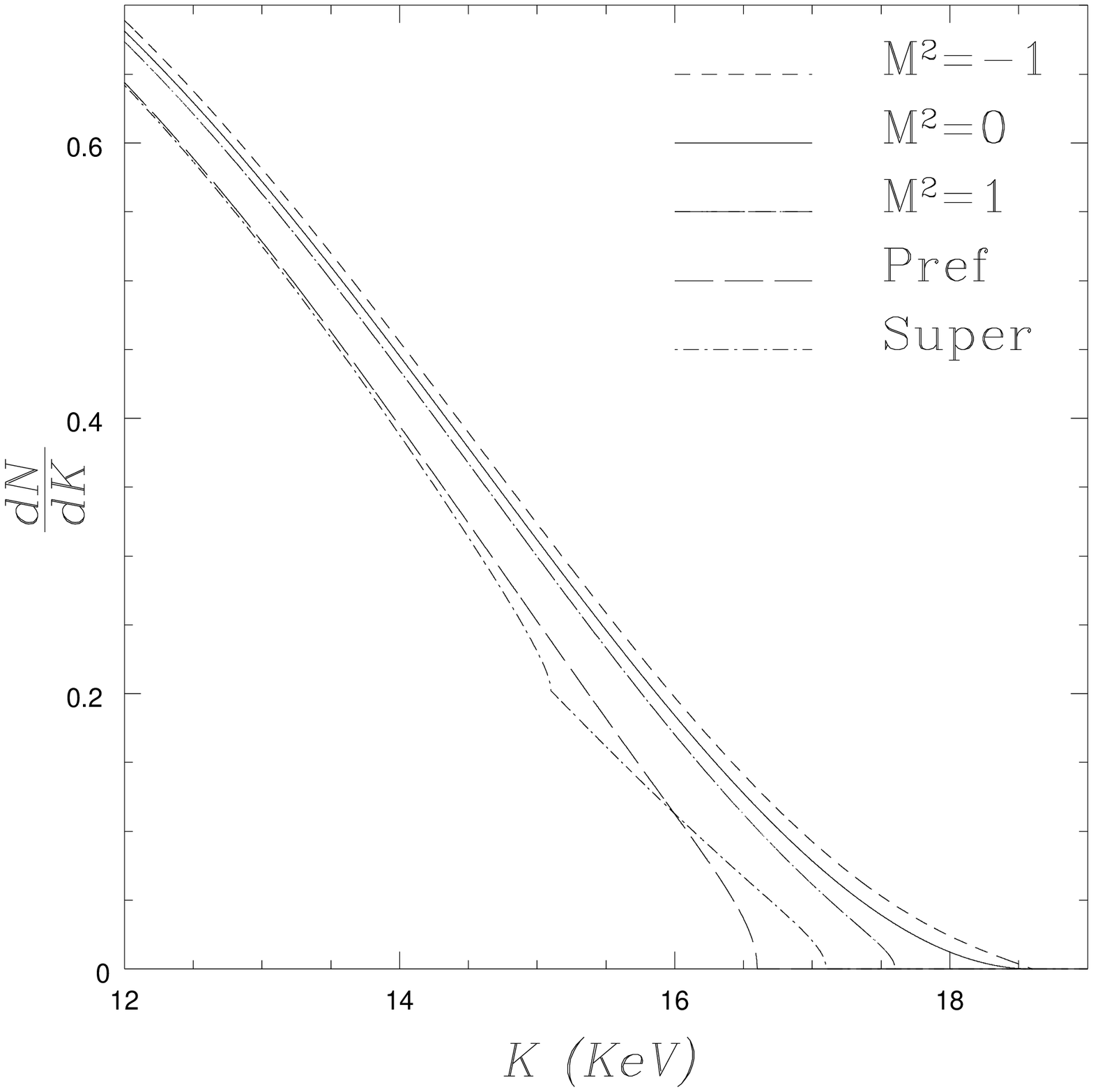, width=0.45\textwidth,clip=} 
\hfill  \epsfig{file=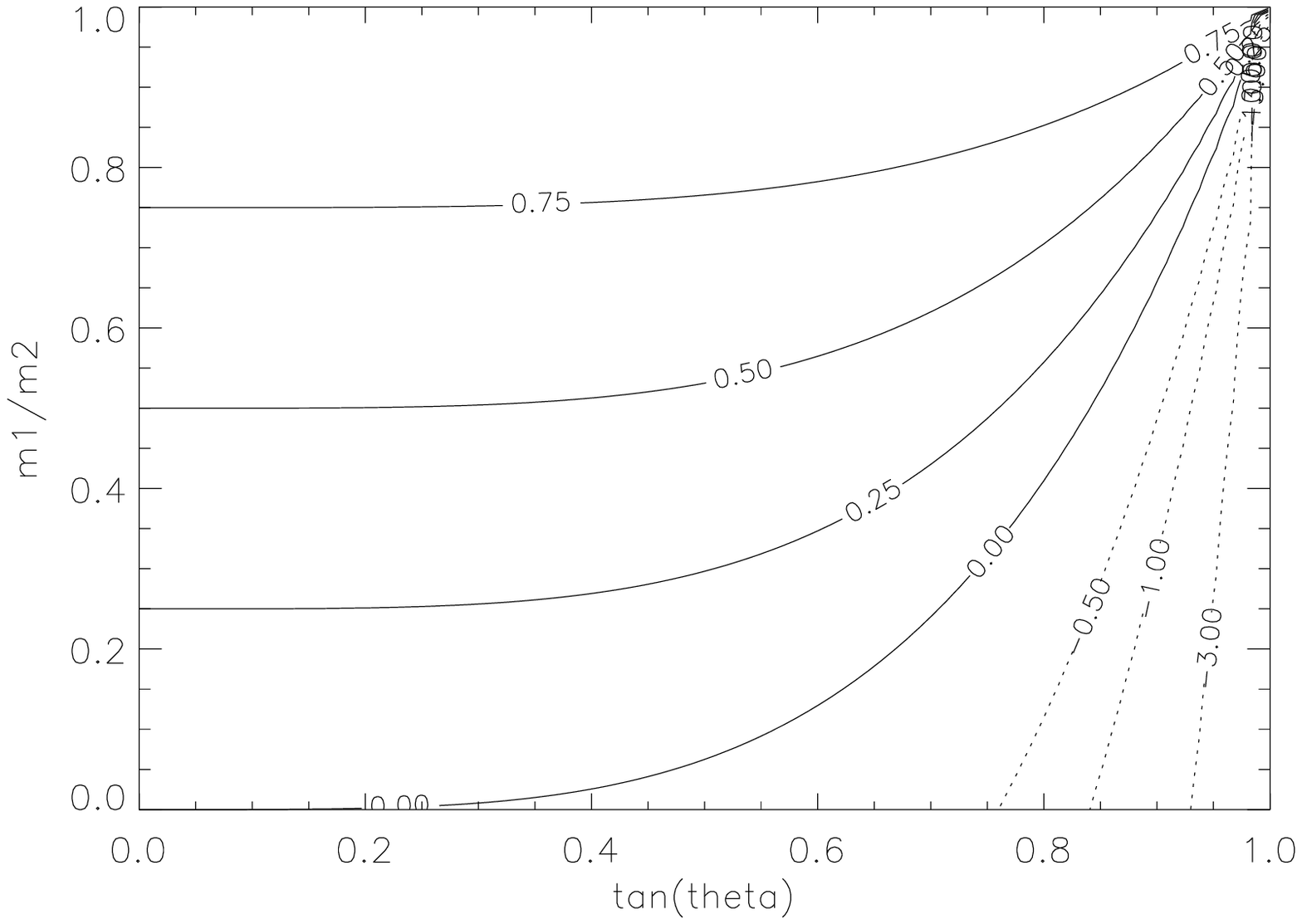, width=0.45\textwidth,clip=}
\caption{{\underline{Left}: Tail of tritium 
$\beta$-decay spectrum, for massless $\nu$ (solid) 
and for LI flavour states (dashed and dot-long-dashed). Also plotted
is the 
preferred frame case. \underline{Right}: Likelihood Contours of $M^2$ (in units of $m_2^2$) upon which $\beta$-decay 
depends (from Ref. [41]).}}
\label{bdecay}
\end{figure}

The interesting feature is that 
these ideas can be tested experimentally, e.g. in $\beta$-decay
experiments:
$N_1 \to  N_2 + e^- + {\bar \nu}_e $, where e.g. $N_1 =~^3{\rm H}$, 
$N_2=~^3{\rm He}$. 

Energy conservation in conventional $\beta$-decay implies: 
$E_{N_1} = E_{N_2} + E + E_e$, where $E$ is the energy of $e$, 
which would unavoidably introduce a preferred
frame. However, in the 
non-linear LI case for flavour states, where 
the use of preferred frame is avoided, this relation 
is modified~\cite{blama}: $E_{N_1} = E_{N_2} + E + E_ef_e(E_e)$. 

These two choices are reflected in different predictions for the 
endpoint of the $\beta$-decay, that is the maximal kinetic energy 
the electron can carry (c.f. figure \ref{bdecay}). 
We refer the interested reader to \cite{blama} for further discussion
on the experimental set up to test these ideas. 

From the point of view of CPTV, which is our main topic of discussion
here, I must mention that in such non-linearly modified Lorentz symmetry
cases it is not clear what form the CPT theorem, if any, takes. 
This is currently 
under investigation~\cite{waldron}. In this sense, the link between CPTV 
and modified flavour-state dispersion relations, and therefore
the interpretation of the associated experiments from this viewpoint,
are issues which are not yet clear, at least to me.

\subsection{CPTV for $\nu$ through QG Decoherence} 

So far, I have discussed the violation of CPT through the violation of 
Lorentz invariance. In this subsection I would like to 
discuss CPTV through decoherence, which is my preferred way of QG-induced
CPTV. As mentioned above, in this case the mater systems are viewed as 
open quantum mechanical or quantum-field theoretic 
systems interacting with a gravitational `environment', consisting
of degrees of freedom inaccessible by low-energy scattering experiments.
The presence of such an environment leads to modified quantum evolution,
which however is 
{\it not necessarily Lorentz Violating}~\cite{mill}. Thus, such an approach
to CPTV should in principle be studied separately, and indeed
it is possible for the 
CPTV decoherence effects to be disentangled experimentally 
from the LV ones, due to the
frame dependence of the latter. 

Before discussing the neutrino phenomenology of this type of CPTV,
it is instructive to mention that 
the currently most sensitive particle physics probes of such a modification from quantum 
mechanical behavior (often called `quantum mechanics violation' 
QMV~\cite{ehns,emn}) are: 
(i) neutral kaons and B-mesons~\cite{ehns,emn}  
and $\phi$-, B-factories (for novel CPT tests for EPR states in such factories
see discussion in~\cite{bernabeu}) 
(ii) ultracold (slow) neutrons in Earth's gravitational field, 
and (iii) Neutrino flavour mixing, which is induced independently of 
masses and mass differences between neutrino species.  In the discussion
below we shall concentrate on this latter probe of QMV. 

Quantum Gravity (QG) may induce oscillations 
between neutrino flavours independently 
of $\nu$-masses~\cite{liu,lisi,benatti,klapdor}. The basic formalism is described by a
QMV evolution for the density matrix of the $\nu$: 
\begin{equation} 
\partial_t \rho = i[\rho, H] +  \delta\H \rho 
\end{equation} 
where~\cite{ehns} 
  \begin{equation}
  {\delta\H}_{\alpha\beta} =\left( \begin{array}{cccc}
 0  &  0 & 0 & 0 \\
 0  &  -2\alpha &  -2\beta  & 0 \\
 0  &  -2\beta  & -2\gamma & 0 \\
 0  &  0 & 0 & 0         \end{array}\right)
\nonumber 
\end{equation}
for energy and lepton number conservation, and 
\begin{equation}
  {\delta\H}_{\alpha\beta} =\left( \begin{array}{cccc}
 0  &  0 & 0 & 0 \\
 0  &  0 & 0 & 0 \\
 0  &  0 & -2\alpha &  -2\beta  \\
 0 & 0  & -2\beta & -2\gamma \end{array}\right)
\nonumber 
\end{equation} 
if energy and lepton number is violated,  
but flavour is conserved (the latter associated formally with the 
$\sigma_1$ Pauli matrix). 

Positivity of $\rho$ requires:
$\alpha, \gamma  > 0,\qquad \alpha\gamma>\beta^2$.
The parameters $\alpha,\beta,\gamma$  violate CP, and CPT in general.   

The relevant oscillation probabilities~\cite{liu} are given: 

{\bf (A)} \underline{For the flavour conserving case}: 

As a simplified example, consider the oscillation 
$\nu_e \to \nu_x$~ ($x=\mu, \tau$ or sterile):

\begin{equation} 
P_{\nu_e \to \nu_x} = \frac{1}{2} - \frac{1}{2}e^{-\gamma L}
{\rm cos}^22\theta_v  - \frac{1}{2}e^{-\alpha L}{\rm sin}^22\theta_v{\rm cos}(\frac{|m_{\nu_1}^2 - m_{\nu_2}^2|}{2E_\nu}L)
\label{prob1}
\end{equation}
Here $L$ is the oscillation length and 
$\theta_v$ is the mixing angle. Note that the mixing angle $\theta_v \ne 0$
if and only if the neutrinos are massless.  

In the mass basis one has:  
$|\nu_e> = {\rm cos}\theta_v |\nu_1> + {\rm sin}\theta_v|\nu_2>,$ 
$|\nu_\mu> = -{\rm sin}\theta_v |\nu_1> + {\rm cos}\theta_v|\nu_2>.$ 

From the above considerations  it is clear that 
there flavour oscillations even in massless case, due to a non-trivial 
QG parameter
$\gamma$, compatible with flavour conserving formalism:
$<\nu_e|\sigma_1|\nu_e>=-<\nu_\mu|\sigma_1|\nu_\mu>= 
2{\rm sin}\theta_v{\rm cos}\theta_v.$

In the above discussion we consider only two flavours. 
For $n$ generations one has: 
$P_{\nu_e \to \nu_x}(t \to \infty)=\frac{1}{n}$. 

{\bf (B)} \underline{For Energy and Lepton number conserving case}:

 Again, we consider a two-flavour example: 
$\nu_e \to \nu_x$~ ($x=\mu, \tau$ or sterile). The relevant oscillation 
probability in this case is calculated to be~\cite{liu}:

\begin{equation} 
P_{\nu_e \to \nu_x} = \frac{1}{2}{\rm sin}^22\theta_v \left(1
- e^{-(\alpha + \gamma) L}
{\rm cos}(\frac{|m_{\nu_1}^2 - m_{\nu_2}^2|}{2E_\nu}L) \right)
\label{prob2}
\end{equation}
where we 
assumed for simplicity, and illustrative purposes, that 
$\alpha,\beta,\gamma \ll \frac{|m_{\nu_1}^2 - m_{\nu_2}^2|}{2E_\nu}$. 

For $n$ generations, the probability reads: 
$P_{\nu_e \to \nu_x}(t \to \infty)=\frac{1}{n}{\rm sin^2}2\theta_v$. 
The reader is invited to contrast this result with case {\bf (A)} above.

One can use 
the results in the cases {\bf (A)} and {\bf (B)} to bound 
experimentally $\xi \equiv \{\alpha,\beta,\gamma$\}. 
At this stage, it is worthy of mentioning that there exist 
two kinds of theoretical estimates/predictions for the order of magnitude
of the parameters $\alpha,\beta,\gamma$:
An optimistic one~\cite{emn}, according to which 
 $\xi \sim \xi_0 (\frac{E}{\rm GeV})^n, 
n=0,2,-1$, and this has a chance of being falsified in 
future experiments, if the effect is there, and a 
pessimistic one, which requires non-trivial masses for $\nu$~\cite{adler}, 
$\xi \sim 
\frac{(\Delta m^2)^2}{E^2 M_{qg}}$, ($M_{qg} \sim M_P \sim 10^{19}$ GeV),
which is much smaller, and probably cannot be accessed by immediate future
neutrino oscillation experiments.

We now mention that in some models of QG-induced decoherence, 
complete positivity  of $\rho(t)$ 
for composite systems, such as $\phi$ or $B$ mesons, may be 
imposed~\cite{benatti} (however, I must note that 
the necessity of this requirement, especially in a QG context
where non-linear effects may be present~\cite{emn}, 
remains to be proven).  
This results in an ideal Markov environment, 
with: $\alpha = \beta = 0, \gamma > 0$.

If this model is 
assumed for $\nu$ oscillations induced by QG decoherence~\cite{lisi}, 
then the following 
phenomenological parametrization can be made:
 $\gamma = \gamma_0 (E/{\rm GeV})^n$, $n=0,2,-1$.
with $E$ the neutrino energy.

\begin{figure}
\centering
 \epsfig{file=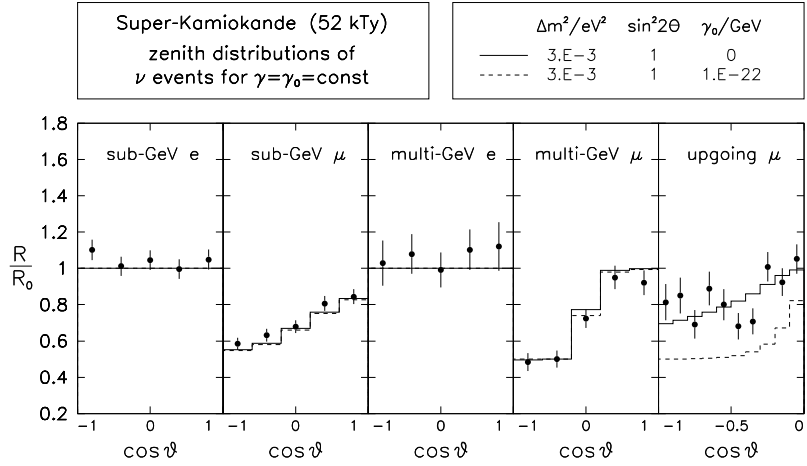, width=0.7\textwidth,clip=}
\caption{Effects of decoherence  
($\gamma = \gamma_0 = {\rm const} \ne 0$)  on the distributions of lepton events as a function of the zenith angle 
$\vartheta$ (from Ref. [43]).}
\label{nudatadec}
\end{figure}

\begin{figure}
\centering
  \epsfig{file=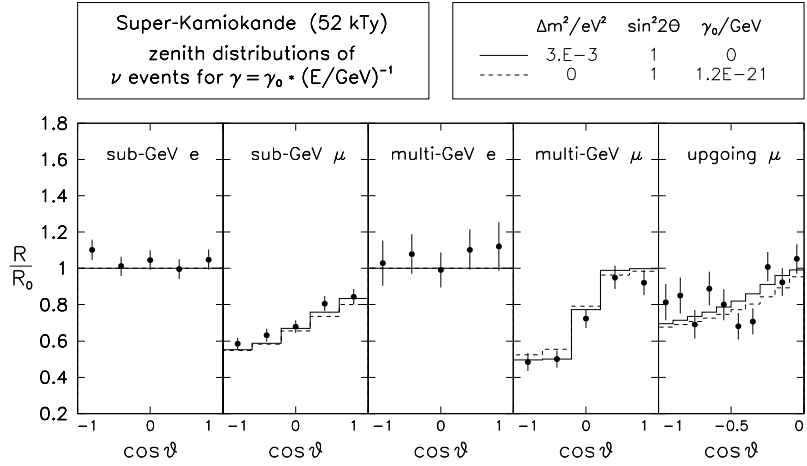, width=0.7\textwidth,clip=} 
\caption{Best-fit scenarios for pure oscillations ($\gamma =0$)
(solid line) and for pure decoherence with  $\gamma \propto 1/E$ 
(dashed line) (from Ref. [43]).}
\label{nudatadec2}
\end{figure}

From Atmospheric $\nu$ data one is led to the following bounds
for the QG-decoherence parameter 
$\gamma$ (c.f. figures \ref{nudatadec},\ref{nudatadec2})~\cite{lisi}:  

(a) $n=0$, $\gamma_0 < 3.5 \times 10^{-23}$ GeV. 

(b) $n=2$, $\gamma_0 < 0.9 \times 10^{-27}$ GeV. 

(c) $n=-1$, $\gamma_0 < 2 \times 10^{-21}$ GeV.  

Especially with respect to case (b) the reader is 
reminded that the 
CPLEAR bound on $\gamma$ for neutral Kaons was 
$\gamma < 10^{-21}$ GeV~\cite{cplear}, i.e. the $\nu$-oscillation experiments
exhibit much higher sensitivity to QG decoherence effects than neutral meson
experiments.

Finally, I note that in \cite{klapdor} it was remarked that 
very stringent bounds on $\alpha,\beta$ and $\gamma$ 
(in the lepton-number violating QG case) 
may be  
imposed by looking 
at oscillations 
of neutrinos from astrophysical sources (supernovae and AGN).
The corresponding bounds
on the $\gamma$ parameter 
from oscillation analysis of neutrinos from supernovae and AGN,
if QG induces such oscillations, are very strong: $\gamma < 10^{-40}$~GeV
from Supernova1987a, using the observed constraint~\cite{sncons} 
on the oscillation 
probability $P_{\nu_e \to \nu_\mu,\tau} < 0.2$, and  
$\gamma < 10^{-42}$~GeV from AGN, 
which exhibit sensitivity  
to order higher than $E^3/M_{qg}^2$, with $M_{qg} \sim M_P \sim 10^{19}$ GeV! 
Of course, the bounds from AGN 
do not correspond to real bounds, 
awaiting the observation of high energy neutrinos 
from such astrophysical sources. 
In \cite{klapdor} bounds have also been derived for the QG decoherence 
parameters by assuming that QG may induce neutrinoless double-beta decay. 
However, using current experimental 
constraints on neutrinoless double-beta decay 
observables~\cite{klapdor2} one arrives at very weak bounds 
for the parameters $\alpha,\beta, \gamma$. 

One also expects stringent bounds on decoherence parameters, 
but also on deformed dispersion relations, if any, for neutrinos,
from future underwater neutrino telescopes, such as ANTARES~\cite{antares},
and NESTOR~\cite{nestor}~\footnote{As far as I understand, but I 
claim no expertise on this issue, the NESTOR experiment has an 
advantage
with respect to detection of very high energy cosmic neutrinos, which may be 
more sensitive probes of such quantum gravity effects.}.

\subsection{CPTV and Departure from Locality for Neutrinos}

\begin{figure}
\centering
  \epsfig{file=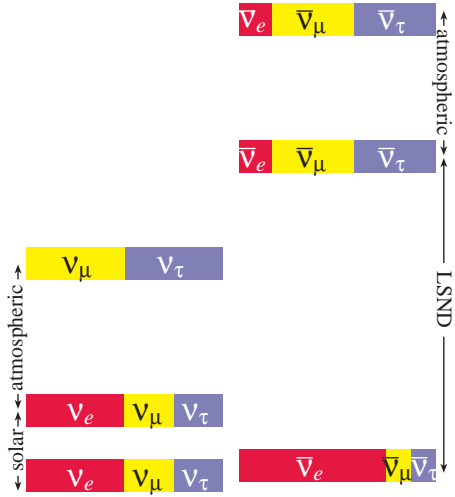, width=0.4\textwidth}
 \caption{The CPTV spectrum 
proposed by Murayama-Yanagida (Ref. [53]). To explain LSND we need 
$m^2_\nu - m^2_{\bar \nu} \sim 0.1~{\rm ev}^{-2} = 10^{-19}~{\rm GeV}^2$.}
\label{nuspectrum}
\end{figure}

As a third way of violating CPT one can relax the requirement 
of {\it locality}. This idea has been pursued in~\cite{barenboim},
in an attempt to present a concrete model for CPT-violation for neutrinos, with CPTV Dirac masses, in an attempt to explain the LSND anomalous 
results~\cite{lsnd}.
In fact the idea of invoking CPTV Dirac mass spectra for neutrinos
in order to account for the LSND results without invoking a sterile neutrino 
is due to the authors 
of \cite{mura} (see figure \ref{nuspectrum}). However no concrete theoretical
model was presented there.

The model lagrangian of \cite{barenboim} reads: 
\begin{eqnarray} 
&& S = \int d^4x {\bar \psi} i\partial_\mu \gamma^\mu 
\psi + \frac{im}{2\pi} \int d^3x dt dt' {\bar \psi}(t) \frac{1}{t - t'}
\psi (t') 
\label{lagrangian}
\end{eqnarray} 
The on shell equations (in momentum space) for the (Dirac) spinors are: 
\begin{eqnarray} 
&&(p_\mu \gamma^\mu - m\epsilon (p_0) )u_\pm (p) =0~, 
\end{eqnarray} 
with $\epsilon (p_0)$ the sign~function, and
\begin{eqnarray} 
&&\psi_+(x) = u_+(p)e^{-ip\cdot x}, \qquad p^2 = m^2, ~p_0 > 0 \nonumber \\
&&\psi_-(x) = u_-(p)e^{-ip\cdot x}, \qquad p^2 = m^2, ~p_0 < 0 
\end{eqnarray} 
Notice that on-shell Lorentz invariance is maintained due to the presence of  
$(\epsilon (p_0)$) but Locality is relaxed.  

As remarked in \cite{greenb}, however, the model of \cite{barenboim},
although respecting Lorentz symmetry on-shell, has 
correlation
functions (which are in general off-shell quantities) that 
do violate Lorentz symmetry, in the sense that they transform 
non covariantly under Lorentz transformations. 

\begin{figure}
\centering
  \epsfig{file=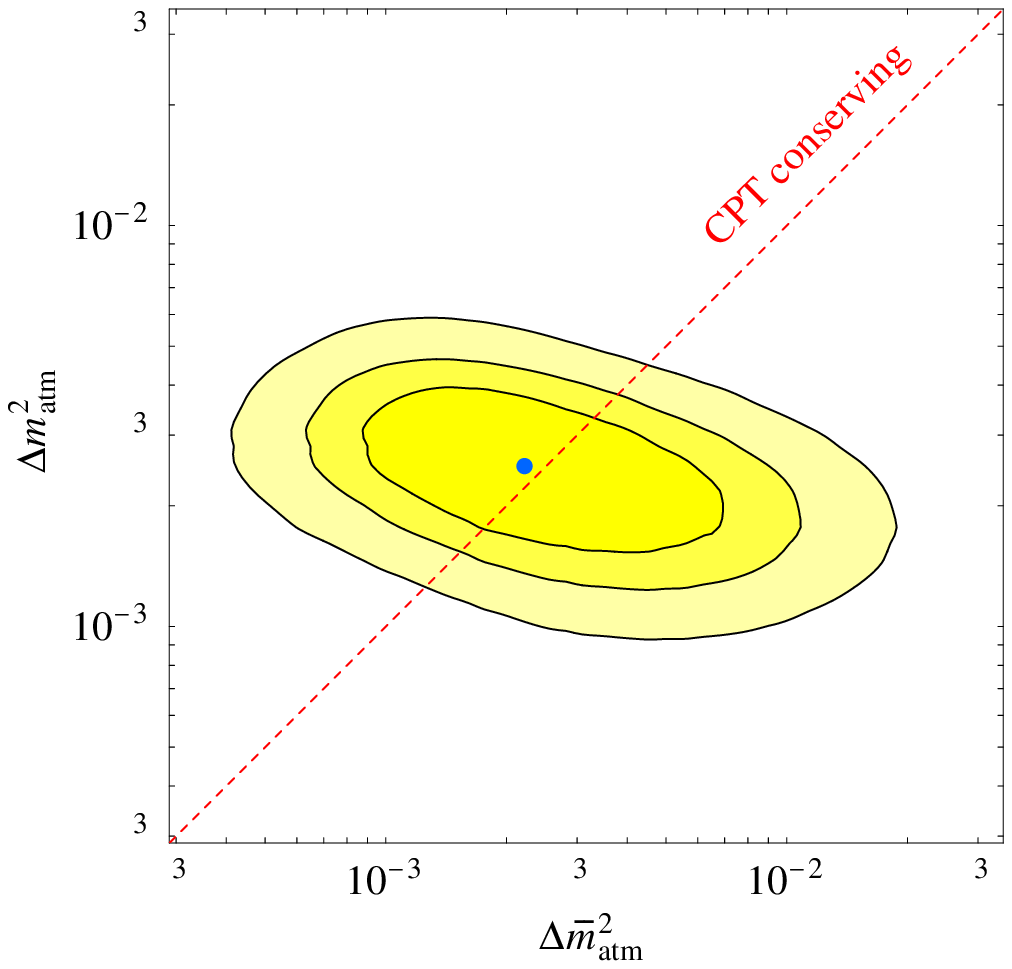, width=0.4\linewidth}
\hfill
\epsfig{file=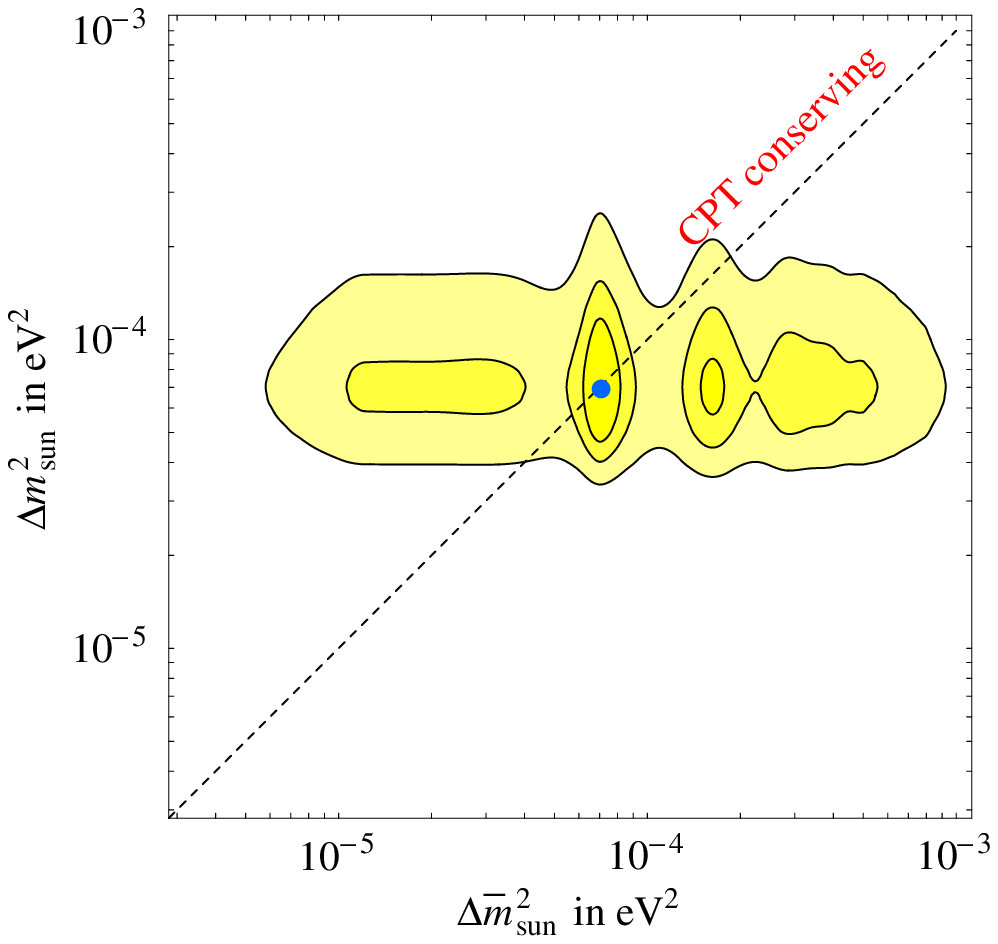, width=0.4\linewidth}
\caption{\underline{Left:} Atmospheric $m_\nu - m_{\overline \nu}$ (68, 90, 99 \%, 2 d.o.f.). \underline{Right:} 
For solar \& reactor data (68, 90, 99 \%, 2 d.o.f.) (from Ref. [54]).}
\label{strumiafig} 
\end{figure}

The two-generation non-local model of \cite{barenboim} 
seems to be marginally disfavoured by the current neutrino data,
as claimed in \cite{strumia} (see figure \ref{strumiafig}).

\begin{table}\small
 \renewcommand{\arraystretch}{1.8}
$$\begin{array}{lc|cccc}
\multicolumn{2}{c|}{\hbox{model \& no.\ of free parameters}} & \Delta\chi^2& \hbox{mainly  incompatible with} &\hbox{main future test}\\ \hline
\multicolumn{2}{c|}{\hbox{ideal fit (no known model)}}& 0 &&?\\
\Delta L = 2\hbox{ decay }\bar\mu\to\bar e \bar\nu_\mu\bar\nu_e & 6 &  12 &\hbox{\sc Karmen}&\hbox{TWIST}  \\
3+1:~\Delta m^2_{\rm sterile} = \Delta m^2_{\rm LSND} & 9  & 6+9?     & \hbox{{\sc  Bugey} + cosmology?}  &\hbox{\sc MiniBoone}\\
\hbox{3 $\nu$ and CPTV~ (no $\Delta \bar m^2_{\rm sun}$)}& 10 &  15  & \hbox{KamLAND} & \hbox{\sc KamLAND}\\
\hbox{3 $\nu$ and CPTV~ (no $\Delta \bar m^2_{\rm atm}$)}& 10 &  25  & \hbox{SK atmospheric} & \hbox{$\bar\nu_\mu$ LBL?}\\
\hbox{normal 3 neutrinos}                             & 5  & 25        &\hbox{LSND}&\hbox{\sc MiniBoone} \\
2+2:~\Delta m^2_{\rm sterile} = \Delta m^2_{\rm sun}  & 9  & 30     & \hbox{SNO}& \hbox{SNO}  \\
2+2:~\Delta m^2_{\rm sterile} = \Delta m^2_{\rm atm}  & 9  & 50    & \hbox{SK atmospheric} & \hbox{$\nu_\mu$ LBL} \\
\end{array}$$ 
  \caption{Interpretations of solar, atmospheric and LSND data, ordered according to
the quality of their global fit.
A $\Delta \chi^2 = n^2$ roughly signals an incompatibility 
at $n$ standard deviations (from Ref.~[54]).}
\end{table}

A summary of data and interpretations of current models, including 
those which entail CPT violation is given in Table 1, taken  from
the first paper in \cite{strumia}. In that paper 
it has also been claimed that the recent 
WMAP~\cite{wmap} data on neutrinos 
seem to disfavour 3 + 1 scenaria
which conserve CPT invariance. In my opinion one has to wait for 
future data from WMAP, before definite conclusions on this issue are reached,
given that the current WMAP data are rather crude in this respect. 
I will not go further into a detailed
discussion of this topic, as such summaries of neutrino data and their 
interpretations 
are covered by other speakers in this conference~\cite{smirnov}.

Before closing this section, I would like to remark 
that most of the theoretical analyses 
for QG-induced CPTV 
in neutrinos have been done in simplified two-flavour oscillation models.
Including all three generations in the formalism 
may lead to differences in the corresponding 
conclusions regarding sensitivity (or conclusions about exclusion) 
of the associated CPTV effects. 
In this respect the measurements of the mixing angle $\theta_{13}$
in the immediate future~\cite{minos}, as a way of detecting 
generic three-flavour effects,  
will be very interesting. In the current phenomenology, CPT invariance
is assumed for the theoretical estimates of this parameter~\cite{lindner}.

\subsection{Four-generation $\nu$ models with CPTV}

\begin{figure}
\centering
  \epsfig{file=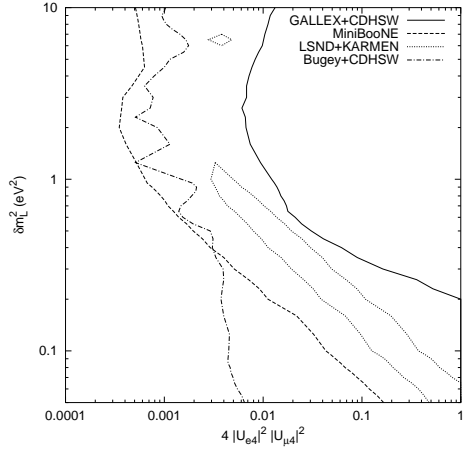, width=0.4\textwidth}
\caption{Upper bound (solid) on the $\nu_\mu \to \nu_e$ oscillation
amplitude $4 |U_{e4}|^2 |U_{\mu4}|^2$ from the GALLEX limit on 
$|U_{e4}|$ and the CDHSW 
limit on $|U_{\mu4}|$ (90\%~C.~L.
results are used in both cases). The dot-dashed line is the 99\% C.~L. 
upper bound from Bugey and CDHSW if $CPT$ is conserved.
Also shown are the expected sensitivity (dashed) of
the MiniBooNE experiment  and, for comparison, the
allowed region (within the dotted lines) 
for $4 |\bar U_{e4}|^2 |\bar U_{\mu4}|^2$ from a
combined analysis of LSND and KARMEN data, both at 
the 90\% C.~L (from Ref. [58]).}
\label{3+1}
\end{figure}

A natural question arises at this point, concerning 
($3 + 1$ or $2 + 2$)  $\nu$ scenaria which violate CPT symmetry. 
This issue has been studied recently in \cite{barger}. 
These authors postulated 
a model for CPTV with four generations for neutrinos 
which leads to different  
flavor mixing between $\nu$, ${\bar \nu}$:
$\nu_a = \sum_{i=1}^{4} U^*_{a i}\nu_i, \qquad 
{\bar \nu}_a = \sum_{i=1}^{4} {\bar U}_{a i}{\bar \nu}_i,$
with $U \ne {\bar U}$ due to CPTV. There are various cases
to be studied:

\begin{figure}
\centering
  \epsfig{file=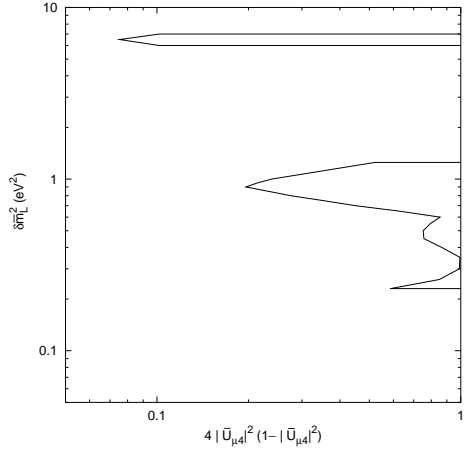, width=0.4\textwidth}
 \caption{Lower bounds on 
$4 |\bar U_{\mu4}|^2 (1 - |\bar U_{\mu4}|^2)$
(the amplitude for atmospheric $\bar\nu_\mu$ survival at the LSND mass
scale) from the Bugey limit on $\bar\nu_e$ disappearance and the
$\bar\nu_\mu \to \bar\nu_e$ oscillation amplitude indicated by LSND and
KARMEN (90\%~C.~L. results are used in both cases) (from Ref. [58]).}
\label{3+1b}
\end{figure}

\begin{figure}
\centering
  \epsfig{file=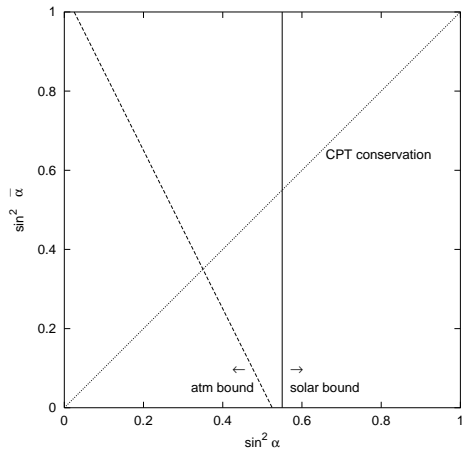, width=0.4\textwidth}
 \caption{Constraints on sterile neutrino mixing angles $\alpha$ and
$\bar\alpha$ from solar (solid) and atmospheric (dashed) data.
The dotted line is the prediction if $CPT$ is conserved (from Ref. [58]).}
\label{2+2}
\end{figure}

\begin{itemize} 
\item{} 3 + 1 models (see figs. \ref{3+1},\ref{3+1b}): one $\nu$ mass well separated from others,
sterile $\nu$ couples only to isolated state. The relevant 
Oscillation probabilities are: $P_{\nu_i \to \nu_i }(|U_{ij}|^2) \ne 
 P_{{\bar \nu}_i \to {\bar \nu}_i }(|{\bar U}_{ij}|^2)  $
  
Experimentally one may bound $|{\bar U}_{e4}|$ and $U_{\mu4}$ 
but there are no tight constraints for $|{\bar U}_{\mu 4}|$, 
$U_{e 4}$. This is to be contrasted with (3 + 1)$\nu$ CPT conserving models 
where $U = {\bar U}$. Hence (3 + 1)$\nu$ + CPTV seems still viable.

 \item{} 2 + 2 models (see fig. \ref{2+2}): sterile $\nu$ couples to solar and 
atmospheric $\nu$ oscillations.  This structure
is only permitted in ${\bar \nu}$ sector. Even with 
CPT Violation, however,  2+2 models are strongly disfavoured by data. 

\end{itemize}

\section{Conclusions} 

From this brief exposition it becomes 
clear, I hope, that CPT Violation may not be an academic issue, 
and indeed it may characterize a theory of quantum gravity. 
As I discussed, neutrino physics provide 
stringent constraints on CPT Violation, which in some cases are much 
stronger  than constraints from neutral meson experiments and factories.
In this sense neutrinos 
may provide a very useful guide
in our quest for a theory of Quantum Gravity. 
For instance, neutrino oscillation experiments provide stringent bounds
on many quantum gravity models entailing Lorentz Invariance Violation.
There are also plenty of low energy nuclear and atomic physics
experiments which yield stringent bounds in models with 
Lorentz (LV) and CPT violation (notice that 
the frame dependence of LV effects
is crucial for such high sensitivities).
It is my firm opinion that neutrino factories, 
when built, will undoubtedly shed light on such important
and fundamental issues and provide definitive answers to many 
questions related with LV models of quantum space time.

But, as I repeatedly stressed, Quantum Gravity may exhibit Lorentz Invariant 
(and hence frame independent) CPTV Decoherence.
 Theoretically the presence of an environment may be consistent
with Lorentz Invariance~\cite{mill}.
 This scenario 
is still compatible with all the existing $\nu$ data, given that 
the parameters of such models are highly model dependent, 
and thus subject at present only to 
constraints by experiment. It is interesting to remark, though, 
that, in cases where quantum gravity induces neutrino oscillations
between flavours or violates lepton number, the sensitivity of experiments
looking for astrophysical neutrinos from extragalactic sources 
may exceed the order of $1/M_P^2$ in the respective figures of merit, and thus
is far more superior than the sensitivities of meson factories 
and nuclear and 
atomic physics experiments, viewed as probes of quantum 
mechanics~\footnote{However, as I remarked previously, 
the reader should be alert to the fact that 
there may be novel CPTV effects unrelated, in principle, 
to LV and locality violations, which are 
associated with modifications of 
EPR correlations. Such effects may be inapplicable to neutrinos, and thus 
testable only in meson 
factories~\cite{bernabeu}.}.  

Clearly much more work, both theoretical and experimental, is needed 
before definite conclusions are reached. Nevertheless, I personally believe 
that research on neutrinos could soon make important contributions to our 
fundamental quest for understanding the quantum structure of space time. 
Neutrino research 
certainly constitutes a very interesting area of fundamental physics, 
which will provide fruitful collaboration between astrophysics
and particle physics, and which,
apart from the exciting possibility 
of non-zero neutrino masses, may still hide even further
surprises waiting to be discovered in the near future.

\section{Acknowledgements}

It is a real pleasure to thank Prof. M. Baldo-Ceolin for 
the invitation and for organizing this very successful and 
thought-stimulating meeting. 
I would also like to acknowledge informative discussions with 
G. Barenboim on Early Universe neutrino physics, 
J. Bernabeu and J. Papavassiliou on CPT phenomenology, and 
G. Tzanakos on the MINOS experiment and prospects for detection of 
generic three-flavour $\nu$ effects.
This work is partly supported by 
the European Union (contract HPRN-CT-2000-00152).

\end{document}